\documentstyle[twoside]{article}

\catcode`\@=11
\long\def\@makefntext#1{
\protect\noindent \hbox to 3.2pt {\hskip-.9pt  
$^{{\eightrm\@thefnmark}}$\hfil}#1\hfill}		

\def\@makefnmark{\hbox to 0pt{$^{\@thefnmark}$\hss}}	
	
\def\ps@myheadings{\let\@mkboth\@gobbletwo
\def\@oddhead{\hbox{}
\rightmark\hfil\eightrm\thepage}   
\def\@oddfoot{}\def\@evenhead{\eightrm\thepage\hfil
\leftmark\hbox{}}\def\@evenfoot{}
\def\sectionmark##1{}\def\subsectionmark##1{}}

\oddsidemargin=\evensidemargin
\newcounter{sectionc}\newcounter{subsectionc}\newcounter{subsubsectionc}
\renewcommand{\section}[1] {\vspace{12pt}\addtocounter{sectionc}{1}
\addtocounter{section}{1} 
\setcounter{subsectionc}{0}\setcounter{subsubsectionc}{0}\noindent 
	{\tenbf\thesectionc. #1}\par\vspace{5pt}}
\renewcommand{\subsection}[1] {\vspace{12pt}\addtocounter{subsectionc}{1} 
	\setcounter{subsubsectionc}{0}\noindent 
	{\bf\thesectionc.\thesubsectionc. {\kern1pt \bfit #1}}\par\vspace{5pt}}
\renewcommand{\subsubsection}[1] {\vspace{12pt}\addtocounter{subsubsectionc}{1}
	\noindent{\tenrm\thesectionc.\thesubsectionc.\thesubsubsectionc.
	{\kern1pt \tenit #1}}\par\vspace{5pt}}

\newcounter{appendixc}
\newcounter{subappendixc}[appendixc]
\newcounter{subsubappendixc}[subappendixc]
\renewcommand{\thesubappendixc}{\Alph{appendixc}.\arabic{subappendixc}}
\renewcommand{\thesubsubappendixc}
	{\Alph{appendixc}.\arabic{subappendixc}.\arabic{subsubappendixc}}

\renewcommand{\appendix}[1] {\vspace{12pt}
        \refstepcounter{appendixc}
        \setcounter{figure}{0}
        \setcounter{table}{0}
        \setcounter{lemma}{0}
        \setcounter{theorem}{0}
        \setcounter{corollary}{0}
        \setcounter{definition}{0}
        \setcounter{equation}{0}
        \renewcommand{\thefigure}{\Alph{appendixc}.\arabic{figure}}
        \renewcommand{\thetable}{\Alph{appendixc}.\arabic{table}}
        \renewcommand{\theappendixc}{\Alph{appendixc}}
        \renewcommand{\thelemma}{\Alph{appendixc}.\arabic{lemma}}
        \renewcommand{\thetheorem}{\Alph{appendixc}.\arabic{theorem}}
        \renewcommand{\thedefinition}{\Alph{appendixc}.\arabic{definition}}
        \renewcommand{\thecorollary}{\Alph{appendixc}.\arabic{corollary}}
        \renewcommand{\theequation}{\Alph{appendixc}.\arabic{equation}}
        \noindent{\tenbf Appendix \theappendixc #1}\par\vspace{5pt}}
\newcommand{\subappendix}[1] {\vspace{12pt}
        \refstepcounter{subappendixc}
        \noindent{\bf Appendix \thesubappendixc. {\kern1pt \bfit #1}}
	\par\vspace{5pt}}
\newcommand{\subsubappendix}[1] {\vspace{12pt}
        \refstepcounter{subsubappendixc}
        \noindent{\rm Appendix \thesubsubappendixc. {\kern1pt \tenit #1}}
	\par\vspace{5pt}}

\newcommand{\textlineskip}{\baselineskip=13pt}
\newcommand{\smalllineskip}{\baselineskip=10pt}

\def\eightcirc{
\begin{picture}(0,0)
\put(4.4,1.8){\circle{6.5}}
\end{picture}}
\def\eightcopyright{\eightcirc\kern2.7pt\hbox{\eightrm c}} 

\newcommand{\copyrightheading}[1]
	{\vspace*{-2.5cm}\smalllineskip{\flushleft
	{\footnotesize International Journal of Modern Physics E, #1}\\
	{\footnotesize $\eightcopyright$\, World Scientific Publishing
	 Company}\\
	 }}

\newcommand{\publisher}[2]{{\begin{center}\footnotesize\smalllineskip 
	Received #1\\
	Revised #2
	\end{center}
	}}

\def\abstracts#1#2#3{{
	\centering{\begin{minipage}{4.5in}\baselineskip=10pt\footnotesize
	\parindent=0pt #1\par 
	\parindent=15pt #2\par
	\parindent=15pt #3
	\end{minipage}}\par}} 

\renewenvironment{thebibliography}[1]
	{\frenchspacing
	 \ninerm\baselineskip=11pt
	 \begin{list}{\arabic{enumi}.}
        {\usecounter{enumi}\setlength{\parsep}{0pt}     
	 \setlength{\leftmargin 12.7pt}{\rightmargin 0pt} 
         \setlength{\itemsep}{0pt} \settowidth
	{\labelwidth}{#1.}\sloppy}}{\end{list}}

\newcounter{itemlistc}
\newcounter{romanlistc}
\newcounter{alphlistc}
\newcounter{arabiclistc}

\newcommand{\fcaption}[1]{
        \refstepcounter{figure}
        \setbox\@tempboxa = \hbox{\footnotesize Fig.~\thefigure. #1}
        \ifdim \wd\@tempboxa > 5in
           {\begin{center}
        \parbox{5in}{\footnotesize\smalllineskip Fig.~\thefigure. #1}
            \end{center}}
        \else
             {\begin{center}
             {\footnotesize Fig.~\thefigure. #1}
              \end{center}}
        \fi}

\newcommand{\tcaption}[1]{
        \refstepcounter{table}
        \setbox\@tempboxa = \hbox{\footnotesize Table~\thetable. #1}
        \ifdim \wd\@tempboxa > 5in
           {\begin{center}
        \parbox{5in}{\footnotesize\smalllineskip Table~\thetable. #1}
            \end{center}}
        \else
             {\begin{center}
             {\footnotesize Table~\thetable. #1}
              \end{center}}
        \fi}

\def\@citex[#1]#2{\if@filesw\immediate\write\@auxout
	{\string\citation{#2}}\fi
\def\@citea{}\@cite{\@for\@citeb:=#2\do
	{\@citea\def\@citea{,}\@ifundefined
	{b@\@citeb}{{\bf ?}\@warning
	{Citation `\@citeb' on page \thepage \space undefined}}
	{\csname b@\@citeb\endcsname}}}{#1}}

\newif\if@cghi
\def\cite{\@cghitrue\@ifnextchar [{\@tempswatrue
	\@citex}{\@tempswafalse\@citex[]}}
\def\citelow{\@cghifalse\@ifnextchar [{\@tempswatrue
	\@citex}{\@tempswafalse\@citex[]}}
\def\@cite#1#2{{$\null^{#1}$\if@tempswa\typeout
	{IJCGA warning: optional citation argument 
	ignored: `#2'} \fi}}

\def\pmb#1{\setbox0=\hbox{#1}
	\kern-.025em\copy0\kern-\wd0
	\kern.05em\copy0\kern-\wd0
	\kern-.025em\raise.0433em\box0}

\def\fnt#1#2{\footnotetext{\kern-.3em
	{$^{\mbox{\scriptsize #1}}$}{#2}}}

\def\fpage#1{\begingroup
\voffset=.3in
\thispagestyle{empty}\begin{table}[b]\centerline{\footnotesize #1}
	\end{table}\endgroup}

\def\runninghead#1#2{\pagestyle{myheadings}
\markboth{{\protect\footnotesize\it{\quad #1}}\hfill}
{\hfill{\protect\footnotesize\it{#2\quad}}}}
\font\tenrm=cmr10
\font\tenit=cmti10 
\font\tenbf=cmbx10
\font\bfit=cmbxti10 at 10pt
\font\ninerm=cmr9
\font\nineit=cmti9
\font\ninebf=cmbx9
\font\eightrm=cmr8

\def\qed{\hbox{${\vcenter{\vbox{			
   \hrule height 0.4pt\hbox{\vrule width 0.4pt height 6pt
   \kern5pt\vrule width 0.4pt}\hrule height 0.4pt}}}$}}

\def\bsc{{\sc a\kern-6.4pt\sc a\kern-6.4pt\sc a}}	
\def\bflatex{\bf L\kern-.30em\raise.3ex\hbox{\bsc}\kern-.14em 
T\kern-.1667em\lower.7ex\hbox{E}\kern-.125em X}

\def\strange{$s$\ }
\def\antistrange{$\bar s$\ }
\def\beqn{\begin{equation}}
\def\eeqn{\end{equation}}
\def\beqna{\begin{eqnarray}}
\def\eeqna{\end{eqnarray}}

\begin{document}


\normalsize\textlineskip
\thispagestyle{empty}
\setcounter{page}{1}

\copyrightheading{}			\runninghead{\today}{\today}

\vspace*{0.88truein}

\fpage{1}
\centerline{\bf NUCLEON STRUCTURE AND}
\centerline{\bf PARITY-VIOLATING ELECTRON SCATTERING}
\vspace*{0.37truein}
\centerline{\footnotesize DOUGLAS H. BECK}
\vspace*{0.015truein}
\centerline{\footnotesize\it Department of Physics, University of Illinois 
at Urbana-Champaign, 1110 West Green Street}
\baselineskip=10pt
\centerline{\footnotesize\it Urbana, Illinois 61801-3080,
USA}
\vspace*{10pt}
\centerline{\footnotesize and}
\vspace*{10pt}
\centerline{\footnotesize BARRY R. HOLSTEIN}
\vspace*{0.015truein}
\centerline{\footnotesize\it Department of Physics-LGRT, University of 
Massachusetts}
\baselineskip=10pt
\centerline{\footnotesize\it Amherst, Massachusetts 01003-4525, USA}
\vspace*{0.225truein}
\publisher{(received date)}{(revised date)}

\vspace*{0.21truein}
\abstracts{We review the area of strange quark contributions to
nucleon structure.  In particular, we focus on current models of
strange quark vector currents in the nucleon and the associated 
parity-violating
elastic electron scattering experiments from which vector and
axial-vector 
currents are extracted.}{}{}


\vspace*{1pt}\textlineskip	
\section{Introduction}		
\vspace*{-0.5pt}
\noindent
A description of the structure of hadrons in terms of quarks and gluons is one 
of the 
challenging problems in modern physics.  Whereas QCD is taken to be the 
correct microscopic basis for such a description, efforts to solve the theory 
directly have 
thus far been unsuccessful for all but the shortest distance scale regimes.  The 
short 
distance scale physics can indeed be calculated from QCD and matches very well 
with 
the body of deep-inelastic scattering\cite{DIS} and e$^+$ e$^-$ annihilation\cite{epem}
data collected over the past 30 years.  In 
the infinite 
momentum frame (where Bjorken scaling is manifest), the nucleon, for example, is 
viewed as primarily composed of its three valence quarks accompanied by ever 
increasing numbers of sea quarks and (especially) gluons as one looks to softer 
and softer 
parts of the wave function.  The gluons, in fact, carry roughly half the nucleon 
momentum overall.  In the same frame, the spin carried by quark
degrees of freedom appears to be only partially 
responsible for the overall spin of the nucleon with the remainder being made up 
from 
quark orbital angular momentum and gluon spin.\cite{SDDIS}

At the longest distance scales (on the order of the size of the hadron),
the picture is at best incomplete.  Models employing effective degrees
of freedom, and which are more or less directly motivated by observation or by QCD,
are the present standards for this regime.  Lattice QCD holds the
promise of being able to calculate many of these low energy observables
in the near future.  It seems inevitable that some synthesis of lattice
results and modeling will be required to lead us to the most economical
approximate descriptions of hadron structure using the appropriate
effective degrees of freedom - a path well worn in other areas of
many-body physics.  Careful measurements of low energy properties will
be necessary to inform the development of these descriptions.

As is the case at the highest energies, the electroweak probe provides a
well understood means of measuring observables at low energies as well.
Measured at low energies, the electromagnetic properties of the nucleon
provided the first indications that it is a composite particle, starting
with the measurement of the proton magnetic moment by Stern, et al.\cite{Stern} in
1933 and continuing with the classic measurements by Hofstadter, et
al.\cite{Hof} of the proton charge and magnetic form factors in the
1950's.  New measurements using both neutrino scattering\cite{bnl,garvey-tayloe} and
parity-violating electron scattering\cite{MusolfReview} provide complementary information
to extend the understanding gained from the earlier studies.

With the unification of descriptions of the electromagnetic and weak
interactions in the early 1970's, it became possible to consider
comparing the electromagnetic and weak observables to extract more
detailed information.\cite{cg,KaMa}  In particular, because the photon and weak gauge
bosons have precisely related couplings to the point-like quarks of the
QCD lagrangian, it is possible to extract structure information according to quark
flavor, as will be discussed in Section~2.  Such determinations are
particularly appropriate for comparison with lattice calculations
because they represent partial, but potentially important, contributions
to the overall low energy structure, providing a more detailed check than do global
observables.

The contribution of strange quarks to the nucleon structure is of
particular interest for developing our understanding, because it is
exclusively part of the quark-antiquark sea.  Further, these pairs
reflect in part the gluon contributions that undoubtedly play an important role,
as they do at
smaller distance scales.  The light (up and down) quark contributions to the sea
may well have important differences relative to those of the strange
quarks - a question that has not yet been thoroughly addressed.
Nevertheless, there is at present no other technique that is directly
sensitive to the quark sea at large distance scales.

That strange quarks make some contribution to nucleon structure is not
in doubt.  Measurement of charm production in deep-inelastic neutrino
scattering provides both the \strange and \antistrange
momentum distributions.  Recently an anaysis by the NuTeV collaboration
has yielded the distinctly nonzero value\cite{nut}
\begin{equation}
{2\int_0^1 dx (s+\bar{s})\over \int_0^1 dx(u+\bar{u}+d+\bar{d})}=
0.42\pm 0.07\pm 0.06
\end{equation}
for $Q^2=16$ GeV$^2$.\footnote{There is at present no
experimentally discernable difference in the \strange and \antistrange
distributions from these studies, but the analysis of the new round of experiments is
continuing.}  We note that the light sea quarks carry about
5\% of the proton momentum in the parton model.  Theoretical work involving 
light front methods
has also suggested a significant strangeness content in the asymptotic 
region.\cite{bm}  In this regard, the direct connection between light front and
low energy manifestations of strangeness is difficult to establish, and 
more work is also needed here.

While it is not clear how to connect these deep-inelastic results to
possible contributions at larger distance scales, the sum rules of
spin-dependent deep-inelastic scattering do yield ground state
properties.  Defining the quark helicity content $\Delta q$ via
\begin{equation}
\Delta q\sigma_\mu=<p,\sigma|\bar{q}\gamma_\mu\gamma_5q|p,\sigma>
\end{equation}
one has the constraint from $\vec{\ell}\vec{N}$ scattering (at first
order in $\alpha_s$) 
\begin{equation}
\int_0^1 dxg_1^p(x)={1\over 2}\left[{4\over 9}\Delta u+{1\over 9}\Delta d
+{1\over 9}\Delta s\right](1-{\alpha_s(q^2)\over \pi})
\end{equation}
When combined with the Bjorken sum rule and its SU(3) generalization
\begin{eqnarray}
\Delta u-\Delta d&=&g_A(0)=F+D\nonumber\\
\Delta u+\Delta d-2\Delta s&=&3F-D
\end{eqnarray}
one finds the solution $\Delta u=0.81,\,\Delta d=-0.42,\, \Delta
s=-0.11$, indicating a small negative value for the strange matrix
element.  This is the simplest analysis possible, however, and there
are ambiguities associated both with the assumption about $SU(3)_f$
symmetry and with gluon contributions that are as yet unresolved.

There exists an alternative probe for this 
matrix element:  neutral current elastic neutrino scattering.
The point here is that the form of the Standard Model axial current is
\begin{equation}
<N|A_\mu^Z|N>={1\over 2}<N|\bar{u}\gamma_\mu\gamma_5u-\bar{d}
\gamma_\mu\gamma_5d-\bar{s}\gamma_\mu\gamma_5s|N>,
\end{equation}
which is purely isovector in the case that the strange matrix element 
vanishes and can therefore be 
exactly predicted from the known charged current axial matrix element.
This experiment was performed at BNL and yielded a result\cite{bnl}
\begin{equation}
\Delta s= -0.15\pm 0.09
\end{equation}
consistent with that found from the deep inelastic sector, but a more
precise value is needed.\cite{garvey-tayloe}

One of the original motivations for work in this area of strange quark
contributions to nucleon structure was the analysis
of the pion-nucleon sigma term\cite{ch} which also yields ground state
(scalar) matrix elements.  The basic idea behind the sigma term analysis is that 
one expects in
the limit of vanishing quark masses that  
the nucleon mass should approach some nonzero value $M_0$ associated with
the gluon content and the $\bar{q}q$ condensate.  
On the other hand, in the real world, with
nonzero quark mass, the nucleon mass is modified to become
\begin{equation}
M_N=M_0+\sigma_s+\sigma
\end{equation}  
where, defining $\hat{m}=(m_u+m_d)/2$, 
\begin{equation}
\sigma_s={1\over 2M_N}<N|m_s\bar{s}s|N>,\quad \sigma={1\over 2M_N}
<N|\hat{m}(\bar{u}u+\bar{d}d)|N>
\end{equation}
are the contributions to the nucleon mass from explicit chiral symmetry
breaking effects ({\it i.e.} non-zero quark mass) involving strange, 
and non-strange quarks
respectively.  One constraint in this regard comes from study of the hyperon
masses, which yields
\begin{eqnarray}
\delta&=&{\hat{m}\over 2M_N}<N|\bar{u}u+\bar{d}d-2\bar{s}s|N>\nonumber\\
&=&{3\over 2}{m_\pi^2\over m_K^2-m_\pi^2}(M_\Xi-M_\Lambda)\simeq 25 MeV
\end{eqnarray}
and increases to about 35 MeV when higher order chiral 
corrections are included.\cite{chiralCorrections}
A second constraint comes from analysis of $\pi N$ scattering, which 
says that $\sigma$ can be extracted directly if an isospin-even combination
of amplitudes could be extrapolated via dispersion relations 
to the (unphysical) Cheng-Dashen point
\begin{equation}
F_\pi^2D^{(+)}(s=M_N^2,t=m_\pi^2)=\sigma
\end{equation}
When this is done the result comes out to
be $\sim$60 MeV, which is lowered to about 45 MeV by higher order chiral
corrections\cite{chicalc}.  If $<N|\bar{s}s|N>=0$, as might be expected from a naive
valence quark picture, then we would expect the value coming from the
hyperon mass limit and that extracted from $\pi N$ scattering to agree.  The
fact that they do not can be explained by postulating the existence of a 
moderate strange quark matrix element
\begin{equation}
f={<N|\bar{s}s|N>\over <N|\bar{u}u+\bar{d}d+\bar{s}s|N>}\simeq 0.1
\end{equation}
implying $M_0\simeq 765$ Mev and $\sigma_s\simeq 130$ MeV, which
seem quite reasonable.  However, recent analyses by Olsson\cite{ol}
and by Pavan\cite{pa} have suggested rather larger values -- $\sim
70-80$ MeV -- for the sigma term, leading to $f\simeq 0.2$, $M_0\simeq
500$ MeV and $\sigma_s\simeq$ 375 MeV, which appear somewhat larger
then one might intutively expect to find.  Hence this
problem as well represents work in progress.\cite{sain}

In contrast to the axial and scalar matrix elements, 
the contribution of the \strange quarks to the vector
currents of the nucleon (ordinary charge and magnetization currents) can
be determined more directly.  The \strange (and the $u$ and $d$) quark
contributions are separated by means of comparison of neutral weak and
electromagnetic elastic scattering measurements at low momentum
transfers.\cite{KaMa,bmck89,be89}  The only assumptions in this case are that the proton and
neutron obey charge symmetry\cite{Mi98} (essentially that under an isospin
rotation $u$ quarks in the proton
become $d$ quarks in the neutron and vice versa) and that the quarks are
point-like, spin 1/2 Dirac particles.  Therefore, such measurements
provide a relatively clean basis from which to describe low energy hadron structure. 

As previously indicated, measurements of neutrino scattering and parity-violating electron
scattering also determine axial-vector current matrix elements.  In
particular, whereas neutrino scattering involves only the axial
current ``seen'' by the Z boson, in parity-violating electron scattering,
the photon can couple to an effective axial current as first pointed out
by Zel'dovich.\cite{Zel}  In the case of the photon, the axial current
coupling involves in addition exchange of a weak neutral boson, either
between the electron and the target, or between quarks in the target
itself.  At present, the interpretation of this measure of the axial
current is beginning to generate significant discussion.

In view of the high level of recent experimental as well as
theoretical work in this area, it is a propitious time to provide an
overview of this field, which is the purpose of the present article.  
The structure of our review is as follows.  In Section 2 the
observables in parity-violating electron scattering are enumerated and Section 3
describes the present status of, and relations among, the model
calculations of these quantities.  In Section 4 a brief summary is given
of experimental techniques and the results to date are presented in
Section 5.  We conclude with a discussion of possible future directions.

\section{Observables}		
\noindent
\subsection{Form factors and quark currents}
\begin{figure}[htbp]
\vspace*{13pt}
\vspace*{1.4truein}		
\includegraphics{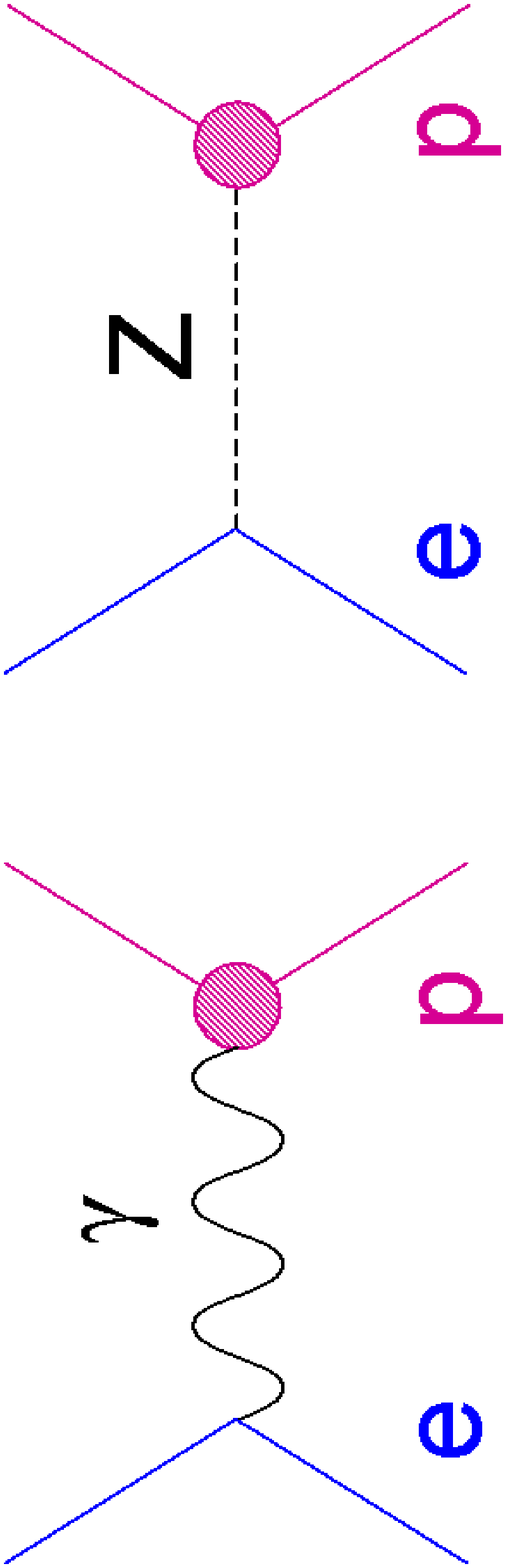}
\vspace*{13pt}
\fcaption{Lowest order Feynman diagrams contributing to electron-nucleon 
scattering.  The electroweak currents of the nucleon are indicated by the shaded 
disks.}
\label{fig:Feynman}
\end{figure}

\noindent
The observables in electroweak electron-nucleon scattering, the overall 
electromagnetic
and neutral weak currents (as indicated in Figure~\ref{fig:Feynman}), may be 
related to the currents 
of the elementary quarks in the nucleon.  Assuming the quarks are point-
like Dirac particles, the form of such currents is
\beqn
J^i_\mu = <N|\sum_{f=flavors}e^i_f \bar q_f \Gamma_\mu q_f |N>
\label{eq:quarkCurrents}
\eeqn
where $i$ denotes the electromagnetic or neutral weak currents and 
$\Gamma_\mu$ is $\gamma_\mu$ for the vector currents and
$\gamma_\mu\gamma_5$ 
for the weak axial current. The electromagnetic charges 
of the quarks, the familiar $2/3$ ($u$, $c$, $t$) and $-1/3$ ($d$, 
$s$, $b$), are related to the neutral weak vector charges
\beqn
e^Z_f = 2 T_{3,f} - 4 e^\gamma_f \sin^2(\theta_W)
\label{eq:weakCharges}
\eeqn 
where $T_{3,f}$ is the weak isospin.\footnote{The definition of the
weak charge is the same as in Ref.~\ref{ref:musolfReview} but, for
example, a factor of 4 larger than Ref.~\ref{ref:Ha84}; and a factor of 2 larger than that of the
Particle Data Group\cite{PDG99}}  When the sum is taken over all 
quark flavors, the above form of the nucleon current is exact for 
point-like Dirac quarks.

Usually, the nucleon current is written as a function of the 
phenomenological elastic form factors (to account for the wave
functions in Eq.~\ref{eq:quarkCurrents}). For example, the electromagnetic 
(vector) current is
\beqn
J^\gamma_\mu = \bar N \left ( F_1(q^2) \gamma_\mu + i \frac{ 
F_2(q^2)}{2M}\sigma_{\mu\nu}q^\nu \right ) N
\label{eq:formFactors}
\eeqn
where now the $N$'s are nucleon spinors and the Dirac ($F_1$) and Pauli 
($F_2$) form factors are normalized to unity and the nucleon anomalous 
magnetic moment, respectively, at $Q^2=0$.  The more common form for 
$F_1$ and $F_2$ is in 
terms of the Sachs (charge and magnetic) form factors
\begin{eqnarray}
G_E &=& F_1 - \frac{Q^2}{4M^2}F_2 \nonumber\\
G_M &=& F_1 + F_2
\end{eqnarray}
The axial current of the nucleon is similarly defined in terms of the 
form factor $G_A$
\beqn
J^Z_{A,\mu} = \bar N G_A(q^2) \gamma_\mu \gamma_5 N.
\eeqn

These form factors can also be expressed in terms of a linear combinations of 
the currents of the different flavors of quarks (as in Eq. 
\ref{eq:quarkCurrents}).  Factoring out the quark charges, the 
electromagnetic and neutral weak vector form factors can be written
\beqna
G^\gamma_{E,M} &=& \frac{2}{3} G^u_{E,M} - \frac{1}{3} (G^d_{E,M} + 
G^s_{E,M}) \nonumber \\
G^Z_{E,M} &=& (1 - \frac{8}{3}\sin^2 \theta_W) G^u_{E,M} + (-
1 + \frac{4}{3}\sin^2 \theta_W) (G^d_{E,M} + G^s_{E,M})
\label{eq:quarkContributionsProton}
\eeqna
Similarly, the neutral weak axial currents of the quarks can be
identified in the overall axial current
\beqn
G_A = G^u_A - (G^d_A + G^s_A).
\label{eq:axialCurrentProton}
\eeqn
These equations illustrate the key point: the electromagnetic and 
neutral weak vector form factors (currents) represent different linear combinations 
of the same matrix elements of contributions from the different flavors 
of quarks.

It should be noted that antiquarks are implicitly included in the
above definitions.  Because they have charges of the opposite sign, quarks
and antiquarks contribute to the matrix elements $G^f_{E,M,A}$ with
opposite signs.  For example, if the spatial distributions of \strange
and \antistrange quarks were the same, their charges would cancel
everywhere and $G_E^s$ would vanish.

Returning to Eqs.~\ref{eq:quarkContributionsProton}, it is
straightforward to solve for the contributions of the three flavors in
the case where one more observable can be found that depends on a
different linear combination of these matrix elements.  By assuming
charge symmetry\cite{Mi98}, the electromagnetic neutron form
factors can be written in terms of the proton matrix
elements.\footnote{The 
assumption is, as mentioned above, that in
exchanging $u$ and $d$ quarks, $\bar u$ and $\bar d$ quarks, and vice
versa, a proton becomes a neutron}  If we
define the $G_E^u$, etc. above as contributions to the {\it
proton} current, then assuming charge symmetry, the corresponding neutron form factors
can be written as 

\beqna G^{n,\gamma}_{E,M} &=& \frac{2}{3} G^d_{E,M}
- \frac{1}{3} (G^u_{E,M} + G^s_{E,M}) \\ 
G^{n,Z}_{E,M} &=&
(1-\frac{8}{3}\sin^2 \theta_W) G^d_{E,M} + (- 1 +
\frac{4}{3}\sin^2 \theta_W) (G^u_{E,M} + G^s_{E,M}) \\ 
G_A^n &=&
G^d_A - (G^u_A + G^s_A)
\label{eq:quarkContributionsNeutron} 
\eeqna

For the record, the {\it vector} current contributions of the three
flavors may be written in terms of the three observables

\beqna
G^{u,p}_{E,M} &=& \left ( 3 - 4 \sin^2\theta_W \right)
G^{p,\gamma}_{E,M} - G^{p,Z}_{E,M} \\ 
G^{d,p}_{E,M} &=& \left ( 2 - 4 \sin^2\theta_W \right)
G^{p,\gamma}_{E,M} + G^{n,\gamma}_{E,M} - G^{p,Z}_{E,M} \\ 
G^{s,p}_{E,M} &=& \left ( 1 - 4 \sin^2\theta_W \right)
G^{p,\gamma}_{E,M} - G^{n,\gamma}_{E,M} - G^{p,Z}_{E,M}
\label{eq:quarkContributions} 
\eeqna

\noindent where again the contributions are shown explicitly to refer to the
proton.

The actual matrix elements corresponding to the nucleon form factors are 
easy to write down in the Breit frame and in the low momentum transfer 
limit.  The matrix elements corresponding to contributions of the 
different quark flavors are the corresponding projections of the overall 
matrix elements. These matrix elements are listed in 
Table~\ref{tab:matrixElements}.
\begin{table}[htbp]
\tcaption{Form factor matrix elements in the Breit frame~\cite{handm} and in the low momentum 
transfer limit.  The operator $J_+$ is that of the normal spherical component of the current.}
\label{tab:matrixElements}
\centerline{\footnotesize\smalllineskip
\begin{tabular}{c c c}\\
\hline
form factor & Breit frame & low momentum transfer limit\\
\hline
$G_E(Q^2)$ & $\frac{1}{2Me} \left < N \right | J_0 \left | N \right >$ &
$\chi_f^\dagger\chi_i(F_1-{Q^2\over 4M^2}F_2)$\\
$G_M(Q^2)$ & $-\frac{1}{2|\vec q|e} \left < N \right | J_+ 
\left | N \right >$ &
$\chi_f^\dagger\vec{\sigma}\chi_i\times\vec{q}{F_1+F_2\over 2M}$\\
$G_A(Q^2)$ &$\left< N\right| \vec{A}\left|N\right>$ & $\chi_f^\dagger\vec{\sigma}\chi_iG_A$\\
\hline\\
\end{tabular}}
\end{table}

Two quantities are often discussed in connection with the behavior of the 
strange quark contributions to these form factors at the lowest momentum 
transfers -- the strangeness (charge) radius, $\rho^s$ and the magnetic moment 
contribution, $\kappa^s$.  They are defined by
\beqna
\lim_{Q^2 \rightarrow 0}G_E^s (Q^2) &=& -\frac{1}{6}Q^2 (\rho^s)^2, \hbox {and} 
\nonumber\\
G_M^s (Q^2=0) &=& \kappa^s
\label{eq:lowq}
\eeqna
A third quantity which is often calculated is a different version of
the strangeness radius---$r_s^2$---measured via the Dirac form factor
\begin{equation}
\lim_{Q^2\rightarrow 0}F_1^s(Q^2)=-{1\over 6}Q^2r_s^2
\end{equation}

\subsection{Higher order effects}
\noindent

\begin{figure}[htbp]
\vspace*{13pt}
\vspace*{1.4truein}		
\includegraphics{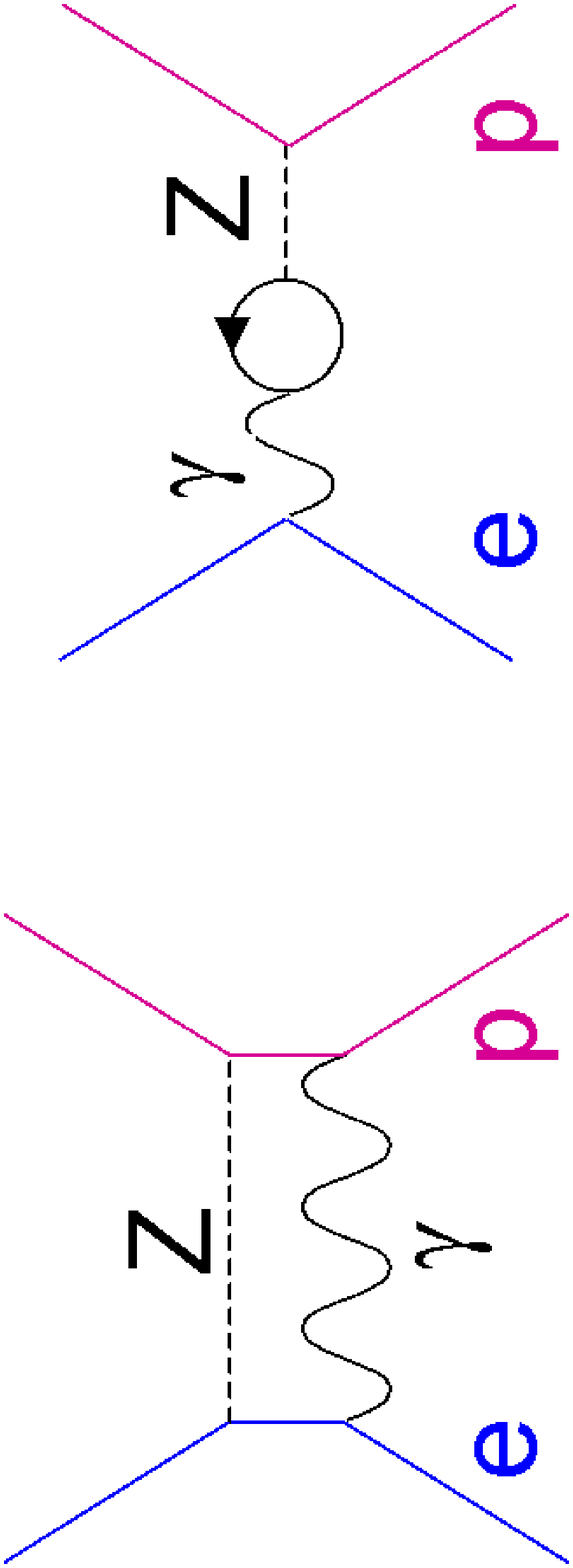}
\vspace*{13pt}
\fcaption{Examples of higher order diagrams that contribute to electroweak 
electron-nucleon scattering.}
\label{fig:higherOrder}
\end{figure}

There are important contributions to electroweak electron-nucleon scattering 
beyond those appearing in Fig.~\ref{fig:Feynman}.  Examples of some of these 
processes are shown in Fig.~\ref{fig:higherOrder}.  
These can be considered radiative corrections to the ordinary weak
$Z^0$ exchange diagram and might be expected therefore to be ${\cal
O}(1\%)$ corrections to the lowest order predictions.  This
assumption is incorrect.  The point is that in order to be 
parity-violating $Z^0$ exchange must involve either 
$$V(e^-)\times A(p)\quad {\rm or} A(e^-)\times V(p)$$
coupling.  However, the weak vector coupling to the electron involves
the factor $1-4\sin^2\theta_W$ and is strongly suppressed.  Thus the
primary sensitivity of a parity-violating electron scattering
experiment is to the case where the weak axial coupling is to the
electron and the weak hadronic interaction is vector in character.
This can be seen explicitly below in Eqs. 29-34.  However, this has
the consequence that, since the piece involving the weak hadronic axial coupling is
suppressed in leading order, the corresponding radiative corrections
to this axial term can be ${\cal O}(\alpha/(1-4\sin^2\theta_W)$ and
quite significant.  This expectation is borne out in detailed
calculations\cite{mjrm}, where corrections to the axial coupling term
involving $G_A^Z$ are found to be as much as 30\%.  However, this
correction is also very uncertain for two reasons
\begin{itemize}   
\item [i)] In diagram \ref{fig:higherOrder}a only 
the component of the box diagram with
a nucleon intermediate state can be reliably computed.  Contributions
from pieces involving excited intermediate state hadrons are unknown.
\item [ii)] In diagram \ref{fig:higherOrder}b wherein
$Z^0$ or $W^\pm$ exchange can occur purely within the hadronic system,
leading to what is sometimes called the anapole moment, the calculations
involve the full complexity of the hadronic state interacting with the 
quark loops.  Although such
effects have been estimated, uncertainties here can be easily 100\%\cite{mjrm}.  
\end{itemize}
Thus the axial current contribution to the asymmetry is not precisely
calculable at the present time and a considerable uncertainly must be
attributed to any such estimate.  This uncertainty acts as an
important systematic background which limits the precision to which
strangeness effects can be measured, and this point will be emphasized below when
the results of the SAMPLE experiment are discussed.  On the other hand, the
anapole related components have a physics interest in their own right, and 
are now beginning to be addressed.

\subsection{Parity-violating electron scattering} 
\noindent 
The neutral weak current of the nucleon can be measured in either
elastic neutrino scattering \cite{garvey-tayloe,Be91} or in 
parity-violating
electron scattering.\cite{KaMa,bmck89,be89}  
Ordinary, unpolarized electron scattering is, of course, dominated
by the electromagnetic currents of the electron and nucleon.  However,
because the weak interaction violates parity (i.e. the weak current
contains both vector and axial-vector contributions), the interference
of the electromagnetic and weak currents violates parity.
Observation of this small effect requires comparison of an
experiment and its mirror image.  In particular, the cross section
contains a pseudoscalar component which will change signs in the
mirror experiment.  In parity-violating electron scattering, the
mirror measurement is made by reversing the (pseudoscalar) beam
helicity, $h = \hat s \cdot \hat p = \pm 1$, where $\hat s$ and $\hat p$ are 
the beam spin and momentum directions, respectively.

It is therefore the asymmetry
\beqn
A^{PV} \equiv \frac {\sigma_+ - \sigma_-}{\sigma_+ + \sigma_-}
\eeqn
that is of interest since the cross section is
\beqn
\sigma \propto \left | {\cal M}^\gamma + {\cal M}^Z \right |^2.
\eeqn
The asymmetry comprises three terms, each reflecting the interference between 
the electromagnetic and neutral weak amplitudes (c.f. Fig.~\ref{fig:Feynman})
\beqn
A_{PV}
= -\frac {G_F Q^2}{4\pi\alpha\sqrt{2}} \frac {A_E + A_M + A_A}{D}
\eeqn
where
\beqna
A_E &=& \epsilon (\theta) G_E^\gamma G^Z_E, \\
A_M &=& \tau G_M^\gamma G_M^Z, \\
A_A &=& - \left ( 1 - 4 \sin^2\theta_W \right ) \sqrt {\tau (1 
+ \tau ) (1 - 
\epsilon(\theta)^2)} G_M^\gamma G_A^Z, \\
D &=& \epsilon (\theta) \left (G_E^\gamma \right )^2 + \tau \left ( 
G_M^\gamma 
\right )^2 \hbox{, and} \\
\epsilon (\theta) &=& \left [ 1 + 2 \left ( 1 + \tau \right ) \tan^2 \theta/2 
\right ] 
^{-1}
\eeqna
where each of the three terms (electric, magnetic and axial) is a
product of electromagnetic and weak form factors as expected.  The
overall scale of the asymmetry is set by the ratio of the neutral weak
and electromagnetic propagators, i.e. 
\beqn
\frac {{\cal M}^Z}{{\cal M}^\gamma}
\sim \frac{Q^2}{(M^Z)^2}.
\eeqn
or about $10^{-4}$ at $Q^2 = 1$ GeV$^2$.  The differing angular
dependence 
associated with these
three terms can be used to separate the contributions experimentally.
Whereas the magnetic term contributes independent of angle, the
electric term vanishes at 180$^\circ$ and the axial term vanishes at
0$^\circ$.  

It is these three terms in the asymmetry that are addressed by the experiments
to be described in Sections 4 and 5.  Before presenting these measurements,
however, we compare and contrast the models used to describe the
strange quark contributions to the nucleon vector currents.

\section{Theoretical Models}
\noindent
As suggested above, a simple three quark picture of the nucleon yields zero for 
strange matrix elements.  However, this is clearly too simplistic.  Indeed,
historically one of the first steps topwards a more realistic, physical decription was to 
consider the nucleon's strangeness content as a 
meson cloud effect -- {\it i.e.} the 
feature that a (``dressed'') nucleon can, for a short time 
consistent with the uncertainty 
principle, transform into a (``bare'') nucleon plus multi-pion state -- within 
which description one can intuitively, but only qualitatively, 
understand the origin of 
the nucleon anomalous moment and charge structure.   In 
the case of the strangeness matrix elements, of course, it is not the 
{\it pion} cloud which is responsible for the effect but rather the 
transformation into states containing strange quarks -- $K\Lambda,\,
K\Sigma,\,\eta N$ -- which yields a nonzero effect.  Estimates based upon
a simple one loop calculation establish the existence of the effect\cite{bh}, 
but, in view of its quantitative failures in predicting the nucleon's 
charge structure and magnetic moment, are certainly expected to be 
unreliable in giving anything other than the order of 
magnitude that one might expect.  

A more contemporary way to describe the origin of nonzero strangeness current
matrix elements is in terms of the quark sea, which can be represented
in terms of the fragmentation of gluons into $q\bar{q}$ pairs.  In this 
regard, the $s\bar{s}$ component is of particular interest since it 
represents the lightest pure sea degree of freedom in the nucleon.  In fact
from the quark sea viewpoint, an intriguing question becomes: since such
sea effects are always present why does the 
simple valence quark model work so well in predicting low energy properties 
of the nucelon?  Possible answers have been given to this question by
Geiger and Isgur who point out that in the adiabatic approximation the 
effect of the virtual pairs is effectively to renormalize the 
string tension, which in 
the quark model is determined phenomenologically and therefore already 
includes such sea effects.\cite{is}  A different answer is given by Kaplan and
Manohar, who suggest that the primary effect of such pairs is to 
renormalize the current quarks of QCD into the constituent quarks used
in phenomenological studies, so that again direct effects are hidden.\cite{KaMa}
Other issues which bear on this question are the origins of the OZI 
rule\cite{is} and the role of gluons in the dynamics of the strange quark 
sea.\cite{qs}
In this paper we will not seek to address such trenchant questions, but 
rather to simply ask how the size of such strangeness current matrix 
elements can be understood from a theoretical perspective.  As we shall
see below, the calculation of quark sea effects within the hadronic
medium is one of the more difficult problems in low energy physics and
a definitive answer will not be possible.  Consequently,
we seek an approach which is model-independent and based on controlled
approximations, but which is at the same time 
effective in establishing contact between theory and the experiment. 

One such general principle is that of (broken) chiral symmetry, which 
is a property of the QCD Lagrangian which presumably underlies all 
interactions.  A way in which one can attempt to insert the strictures of
chiral invariance is through the use of chiral perturbation theory\cite{gl}, 
specifically through its heavy baryon version.\cite{bkm}  The basic idea is 
that in a world in which the light ($u,d,s$) current quark masses vanish, the
QCD Lagrangian\footnote{Here the covariant derivative is
\begin{equation}
i D_{\mu}=i\partial_{\mu}-gA_\mu^a {\lambda^a \over 2} \, ,
\end{equation}
where $\lambda^a$ (with $a=1,\ldots,8$) are the SU(3) Gell-Mann matrices,
operating in color space, and the color-field tensor is defined by
\begin{equation}
G_{\mu\nu}=\partial_\mu  A_\nu -  \partial_\nu  A_\mu -
g [A_\mu,A_\nu]  \, ,
\end{equation} }
\begin{equation}
{\cal L}_{\mbox{\tiny QCD}}=\bar{q}(i  {\not\!\! D} - m )q-
{1\over 2} {\rm tr} \; G_{\mu\nu}G^{\mu\nu} \, .
\end{equation}
would possess an exact $SU(3)_L\times SU(3)_R$ symmetry under independent
left- and right-handed rotations of the light quarks
\begin{equation}
q_L  \rightarrow \exp (i \sum_j \lambda_j\alpha_j)
q_L,\qquad
q_R  \rightarrow \exp (i\sum_j \lambda_j \beta_j)
q_R
\end{equation}
where here left and right refer to actions of the chirality projectors
$P_{L,R}=(1\pm\gamma_5)/2$.  The spontaneous symmetry breaking in the axial
sector which takes $SU(3)L\times SU(3)_R\rightarrow SU(3)_V$ implies via 
Goldstone's theorem\cite{go} the existence of eight massless 
pseudoscalar particles,
which we identify with the $J^P=0^-$ octet and an axial condensate, which
we associate with the pion decay constant $F_\pi$=92.4 MeV.  
(In the real world, 
of course, the current quark masses are small but nonzero, and so these 
bosons are no longer massless but merely lighter than their hadronic siblings.)
These bosons are described via the field
\begin{equation}
U\equiv\exp(i\sum_j\lambda_j\phi_j/F_\pi)
\end{equation} 
where $\lambda_j$ are the Gell-Mann matrices and $\phi_j$ are the pseudoscalar 
fields.  The lowest order chiral effective Lagrangian is then
\begin{equation}
 {\cal L}_2={F_\pi^2 \over 4} \mbox{Tr} (\partial_{\mu}U \partial^{\mu}
 U^{\dagger})+{B_0\over 4} F_\pi^2 \mbox{Tr} m(U+U^{\dagger})\,.\label{eq:abc}
\end{equation}   
where $m$ is the quark mass matrix and $B_0$ is a phenomenological constant
related to the $\bar{q}q$ condensate.  To order $\phi^2$ Eq. \ref{eq:abc}
reproduces the free meson Lagrangian and the Gell-Mann-Okubo
relation\cite{gmo}
\begin{equation}
3m_{\eta}^2 +m_{\pi}^2 - 4m_K^2 =0 
\end{equation}
while at order $\phi^4$ it yields the Weinberg $\pi\pi$ scattering 
lengths\cite{wb}
\begin{equation}
a_0^0={7m_\pi\over 32\pi F_\pi^2},\quad a_0^2=-{m_\pi\over 16\pi F_\pi^2}
\end{equation}
Such effective interaction calculations in the meson sector have been 
developed to a high degree over the past fifteen or so years, including the
inclusion of loop contributions, in order to preserve crossing symmetry and
unitarity.  When such loop corrections are included one must augment the
effective lagrangian to include ``four-derivative'' terms with arbitrary
coefficients which must be fixed from experiment.  This program, called
chiral perturbation theory, has been enormously successful in describing
low energy interactions in the meson sector.\cite{cpt}  However, 
in order to discuss nucleons, we must extend it to consider baryons.

Writing
down the lowest order such chiral Lagrangian at the SU(2) level is
straightforward -- 
\begin{equation}
{\cal L}_{\pi N}=\bar{N}(i\not\!\!{D}-m_N+{g_A\over 2}\rlap /{u}\gamma_5)N
\end{equation}
where $g_A$ is the usual nucleon axial coupling in the chiral limit, the
covariant derivative $D_\mu=\partial_\mu+\Gamma_\mu$ is given by
\begin{equation}
\Gamma_\mu={1\over 2}[u^\dagger,\partial_\mu u]-{i\over 2}u^\dagger
(V_\mu+A_\mu)u-{i\over 2}u(V_\mu-A_\mu)u^\dagger ,
\end{equation}
and $u_\mu$ represents the axial structure
\begin{equation}
u_\mu=iu^\dagger\nabla_\mu Uu^\dagger
\end{equation}
The quantities $V_\mu,\,A_\mu$ represent external (non-dynamical) vector,
axial-vector fields.
Expanding to lowest order we find
\begin{eqnarray}
{\cal L}_{\pi N}&=&\bar{N}(i\rlap /{\partial}-m_N)N+g_A
\bar{N}\gamma^\mu\gamma_5{1\over 2}\vec{\tau}N\cdot({i\over 
F_\pi}\partial_\mu\vec{\pi}
+2\vec{A}_\mu)\nonumber\\
&-&{1\over 4F_\pi^2}\bar{N}\gamma^\mu\vec{\tau}N\cdot\vec{\pi}\times
\partial_\mu\vec{\pi}+\ldots
\end{eqnarray}
which yields the Goldberger-Treiman relation,
connecting strong and axial couplings of the nucleon system\cite{gt}
\begin{equation}
F_\pi g_{\pi NN}=m_N g_A
\end{equation}
Extension to SU(3) gives additional successful predictions -- the baryon
Gell-Mann-Okubo formula as well as the generalized Goldberger-Treiman
relation.  However, difficulties arise when one attempts to include
higher order corrections to this formalism.  The difference from the
Goldstone case is that there now exist {\it three} dimensionful
parameters -- $m_N$, $m_\pi$ and $F_\pi$ -- in the problem rather than
just $m_\pi$ and $F_\pi$.  Thus loop effects can be of order 
$(m_N/4\pi F_\pi)^2
\sim 1$ and we no longer have a reliable perturbative scheme.  A
consistent power counting mechanism {\it can} be constructed provided that we
eliminate the nucleon mass from the leading order Lagrangian, as done in
the procedure of Foldy and Wouthuysen\cite{fw}, which involves diagonalization
of the particle and antiparticle components of the interaction.
Equivalently one generates an expansion in powers of inverse nucleon mass
by writing nucleon four-momenta as\cite{jm}
\begin{equation}
p_\mu=Mv_\mu+k_\mu
\end{equation}
where $v_\mu$ is the four-velocity and satisfies $v^2=1$, while $k_\mu$
is a ``small'' off-shell momentum, with $v\cdot k<< M$.  One can construct
eigenstates of the projection operators $P_\pm = {1\over 2}(1\pm
\rlap /{v})$, which in the rest frame select upper, lower
components of the Dirac wavefunction, so that\cite{bkkm}
\begin{equation}
\psi=e^{-iMv\cdot x}(H_v+h_v)
\end{equation}
where
\begin{equation}
H_v=P_+\psi,\qquad h_v=P_-\psi
\end{equation}
The $\pi N$ Lagrangian can then be written in terms of $H,h$ as
\begin{equation}
{\cal L}_{\pi N}=\bar{H}_v{\cal A}H_v+\bar{h}_v{\cal B}H_v+
\bar{H}_v\gamma_0{\cal B}^\dagger\gamma_0h_v-\bar{h}_v{\cal C}h_v\label{eq:hh}
\end{equation}
where the operators ${\cal A}, {\cal B},{\cal C}$ have the low energy
expansions
\begin{eqnarray}
{\cal A}&=&iv\cdot D+g_A u\cdot S +\ldots\nonumber\\
{\cal B}&=&i\not\!\!{D}^\perp-{1\over 2}g_A v\cdot u\gamma_5+\ldots\nonumber\\
{\cal C}&=&2M+iv\cdot D+g_A u\cdot S+\ldots
\end{eqnarray}
Here $D_\mu^\perp=(g_{\mu\nu}-v_\mu v_\nu)D^\nu$ is the transverse component
of the covariant derivative and $S_\mu={i\over 2}\gamma_5
\sigma_{\mu\nu}v^\nu$ is the Pauli-Lubanski spin vector and satisfies
\begin{equation}
S\cdot v=0,\quad S^2=-{3\over 4},\quad\{S_\mu,S_\nu\}={1\over 2}(v_\mu v_\nu-
g_{\mu\nu}),\quad [S_\mu,S_\nu]=i\epsilon_{\mu\nu\alpha\beta}v^\alpha S^\beta
\end{equation}
We observe that the two components H, h are coupled in Eq. \ref{eq:hh}.
However, the system may be diagonalized by use of 
the field transformation
\begin{equation}
h'=h-{\cal C}^{-1}{\cal B}H
\end{equation}
in which case the Lagrangian becomes
\begin{equation}
{\cal L}_{\pi N}=\bar{H}_v({\cal A}+(\gamma_0{\cal B}^\dagger
\gamma_0){\cal C}^{-1}{\cal B})H_v-\bar{h}'_v{\cal C}h'_v
\end{equation}
The piece of the Lagrangian involving $H$ contains the mass only in 
the operator
${\cal C}^{-1}$ and is the effective Lagrangian that we seek.  The remaining
piece involving $h'_v$ can be thrown away, as it does not couple to the
$H_v$ physics.  (In path integral language we simply integrate out this
component yielding an uninteresting overall constant.)

Of course, for applications involving strangeness, we must generalize the
above to SU(3) and the lowest order heavy baryon chiral Lagrangian
becomes
\begin{equation}
{\cal L}=i{\rm tr}\left(\bar{B}v\cdot DB\right)+2{\rm tr}\left(
\bar{B}S^\mu\left[D\{A_\mu,B\}+F[A_\mu,B]\right]\right)
\end{equation} 
where $D,F$ are the usual SU(3) axial couplings with $D=F=g_A$.
When loops are included one can be concerned about the convergence
of such a chiral series, but the above formalism provides at least a 
useful framework for
our discussions.\footnote{Indeed, as we shall discuss below, simple
SU(3) loop calculations in heavy baryon chiral perturbation theory generate
nonanalytic terms which are large and phenomenologically problematic.}
The inclusion of loops also requires 
the insertion of
phenomenological counterterms in order to absorb divergences.  In our case 
we shall represent the strangeness current in terms the electromagnetic and
baryon number currents -- 
\begin{equation}
J_\mu^{em}(I=1)=\bar{q}{\lambda_3\over 2}\gamma_\mu q,\quad
J_\mu^{em}(I=0)=\bar{q}{\lambda_8\over 2\sqrt{3}}\gamma_\mu q,\quad
J_\mu^B={1\over 3}\bar{q}\gamma_\mu q
\end{equation}
via
\begin{equation}
\bar{s}\gamma_\mu s=J_\mu^B-2J_\mu^{em}(I=0)
\end{equation}
and the appropriate counterterms are given by\cite{mi}
\begin{eqnarray}
\Delta{\cal L}^{I=1}_{em}&=&{e\over \Lambda_\chi}\epsilon_{\mu\nu\alpha\beta}
v^\alpha\left\{b_+{\rm tr}(\bar{B}S^\beta\{\lambda_3,B\})
+b_-{\rm tr}(\bar{B}S^\beta[\lambda_3,B])\right\}F^{\mu\nu}\nonumber\\
&-&{e\over \Lambda_\chi^2}\left\{c_+{\rm tr}\bar{B}\{\lambda_3,B\}
+c_-{\rm tr}\bar{B}[\lambda_3,B]\right\}v_\mu\partial_\nu F^{\mu\nu}\nonumber\\
\Delta{\cal L}^{I=0}_{em}&=&{e\over \sqrt{3}\Lambda_\chi}
\epsilon_{\mu\nu\alpha\beta}
v^\alpha\left\{b_+{\rm tr}(\bar{B}S^\beta\{\lambda_8,B\})
+b_-{\rm tr}(\bar{B}S^\beta[\lambda_8,B])\right\}F^{\mu\nu}\nonumber\\
&-&{e\over \sqrt{3}\Lambda_\chi^2}\left\{c_+{\rm tr}\bar{B}\{\lambda_8,B\}
+c_-{\rm tr}\bar{B}[\lambda_8,B]\right\}v_\mu\partial_\nu F^{\mu\nu}\nonumber\\
\Delta{\cal L}_B&=&{b_0\over \Lambda_\chi}\epsilon_{\mu\nu\alpha\beta}
v^\alpha{\rm tr}\bar{B}S^\beta BZ^{\mu\nu}-{c_0\over \Lambda_\chi^2}
{\rm tr}\bar{B}Bv_\mu\partial_\nu Z^{\mu\nu}
\end{eqnarray}
Here $\Lambda_\chi=4\pi F_\pi\sim 1$ GeV is the chiral scale parameter and
is inserted in order that the parameters $b_i,c_i$ should be ${\cal O}(1)$
while $F^{\mu\nu},Z^{\mu\nu}$ are field tensors for external photon and
Z-boson fields respectively.

One can then represent the experimental values of strangeness anomalous
moments and charge radii (see Eqs.~\ref{eq:lowq}) in terms of the sum of 
counterterm and loop 
contributions as 
\begin{eqnarray}
\kappa^a&=&\kappa^a_{loop}+\left({2M_B\over \Lambda_\chi }\right)b^a\nonumber\\
\rho^a&=&\rho^a_{loop}-\left({2M_B\over \Lambda_\chi }\right)^2c^a
\end{eqnarray}
where the one-loop contributions to the strangeness form factors are\cite{mi} 
\begin{eqnarray}
\kappa^s_{loop}&=&2\pi{M_Nm_K\over \Lambda_\chi^2}\left[{1\over 6}
(3F+D)^2+{3\over 2}(D-F)^2\right]\nonumber\\
\rho^s_{loop}&=&\left({M_N\over \Lambda_\chi }\right)^2
\left[{1\over \epsilon}
-\gamma-{\ln {m_K^2\over 4\pi \mu^2}}\right]\nonumber\\
&\ &\left\{1+{5\over 3}
\left[{1\over 6}
(3F+D)^2+{3\over 2}(D-F)^2\right]\right\} \label{eq:lo}
\end{eqnarray}
Here dimensional regularization has been used with $\epsilon=(4-d)/2$.  
Similar forms obtain for the electromagnetic loop contributions and 
in this way, using experimental values for the neutron and proton form
factors one can obtain values for the counterterms $c_\pm,b_\pm$ -- $b_+
\simeq 1.4,\,b_-\simeq 0.9,\,c_+\simeq -1.9,\,c_-\simeq 0.9$.  (Note
that these values are ${\cal O}(1)$, as expected from chiral counting
arguments.)  The values which we need in order to evaluate the 
strangeness terms are\footnote{Note here that the term $c_0$ contains the
appropriate divergence needed to cancel that given in the loop calculation.}
\begin{eqnarray}
b^s&=&b_0-2[b_--{1\over 3}b_+]\nonumber\\
c^s&=&c_0-2[c_--{1\over 3}c_+]
\end{eqnarray}
so that if values for $c_0,b_0$ were known corresponding to those for
$c_\pm,b_\pm$ our problem would be solved.  Unfortunately this is not the
case.  Indeed, chiral perturbation theory relies on {\it experiment} to
determine those quantities whose value is not required by chiral
invariance so that in our case we require experimental values of the
strangeness anomalous moment and charge radius in order to determine
the missing counterterms -- the theory is nonpredictive.
As discussed below, it is possible to get around this difficulty by
assuming a model, such as vector dominance, but this is beyond the
basic strictures required by chiral invariance alone.  An interesting 
exception to this chiral non-predictability arises, however, if one considers
the $q^2$-dependence of the strange magnetic form factor, as pointed
out by Hemmert, Meissner and Steininger.\cite{hms}  
In this case, although the strange magnetic moment itself is 
not predicted, the momentum tranfer dependence of the magnetic form factor 
{\it is} determined by the kaon loop diagram so that
one has the model-independent prediction
\begin{eqnarray}
F_2^s(q^2)-F_2^s(0)&=&{M_Nm_K\over 24\pi F_\pi^2}(5D^2-6DF+9F^2)
f(q^2)\nonumber\\
\hbox{where} && f(q^2)=-{1\over 2}+{1+{q^2\over 4m_K^2}\over 2\sqrt{q^2\over 4m_K^2}}
\tan^{-1}\left(\sqrt{q^2\over 4m_K^2}\right)
\end{eqnarray}
One can use this expression in order to extrapolate from negative $q^2$,
where experiments are done to $q^2=0$, which is 
where most theoretical calculations are performed.  In the case of the
SAMPLE
experiment (see Section 5), this produces a shift of about 0.2 nucleon magnetons.

\subsection{Vector dominance}

Despite the inability of chiral methods to provide predictions for 
strangeness vector currents, the above formalism does provide a
useful basis from which to take our calculation further.  What we 
require are (of necessity 
model-dependent) methods by which to estimate the size of the chiral 
counterterms.  In 
principle lattice methods can be used in this regard, and work 
in this area is in progress.\cite{lat}  However, reliable
estimation of subtle quark sea effects requires proper inclusion of
quark loop contributions, while published calculations involve
the quenched approximation, wherein quark-antiquark loop effects are 
neglected.\cite{sl}  Thus we require alternative means, which implies
resorting to various hadronic models. 
An approach which has proved useful in the chiral mesonic sector is
to estimate the size of counterterms via saturation by t-channel vector
and axial-vector meson exchange contributions.\cite{ec}  
In this way five of the ten 
${\cal O}(p^4)$ Gasser-Leutwyler counterterms evaluated at a 
renormalization scale $\mu=m_\rho$ are reproduced rather successfully.
In particular, the pion charge radius is predicted to be given predominantly
in terms rho-meson exchange with only a small ($<10\%$) loop contribution.
This, of course, is simply the vector dominance assumption, which is 
known to work well for this quantity.  It then seems reasonable to attempt
a similar calculation in the realm of baryons.  For this purpose, it is
useful to emply a tensor representation of the vector meson fields, as this
automatically builds in proper asymptotic properties.\cite{mi}  In particular, 
defining a VNN effective Lagrangian in terms of couplings $G_T,G_V$ 
via\cite{mi}
\begin{equation}
{\cal L}_{VNN}=2G_T\epsilon_{\mu\nu\alpha\beta}v^\alpha\bar{B}
S^\beta BV^{\mu\nu}+{G_V\over \Lambda_\chi}\bar{B}Bv_\mu D_\nu V_{\mu\nu}
\end{equation}    
as well as a gauge invariant vector-meson photon coupling 
\begin{equation}
{\cal L}_{V\gamma}={eF_V\Lambda_\chi\over \sqrt{2}}F_{\mu\nu}V^{\mu\nu}
\end{equation}
one finds contributions to the charge and magnetic form factors
\begin{eqnarray}
F_1(q^2)&=&\sqrt{2}G_VF_V{Q^2\over m_V^2-q^2}\nonumber\\
F_2(q^2)&=&4\sqrt{2}G_TF_V{M_N\Lambda_\chi\over m_V^2}{m_V^2\over m_V^2-q^2}
\end{eqnarray}
One can then identify the counterterm predictions in terms of
\begin{eqnarray}
b&=&2\sqrt{2}G_TF_V\left({\Lambda_\chi\over m_V}\right)^2\nonumber\\
c&=&\sqrt{2}G_VF_V\left({\Lambda_\chi\over m_V}\right)^2
\end{eqnarray}
where we have for simplicity omitted the isospin designations.  For
the nucleons, the experimental electromagnetic form factors together with
data on the leptonic decay of the vector mesons allows a determination
of the couplings $G_V,G_T,F_V$.  However, lacking knowledge of the 
couplings $F_V$ associated with the strange matrix elements $<0|\bar{s}
\gamma_\mu s|V>$ we have no predictive power in the strange sector.

Nevertheless, progress can be made by invoking the vector meson 
wavefunction quark content.\cite{ja}  The idea here is to 
write down a dispersion
relation approach to these form factors and to assume vector meson dominance
\begin{eqnarray}
F_1^a(q^2)&=&F_1^a+\sum_V{q^2a_V^a\over m_V^2-q^2}\nonumber\\
F_2^a(q^2)&=&\sum_V{m_V^2b_V^a\over m_V^2-q^2}
\end{eqnarray}
Here the isoscalar residues at the vector meson poles have been determined by 
H\"{o}hler et al.\cite{ho} and independently by Mergell et al.\cite{mer} 
in terms of a fit involving $\omega,\phi$ and one higher mass
vector meson $V'$.  Clearly at least two such poles are 
required in order
to obtain the observed dipole dependence while a third pole was used in order
to achieve a reasonable $\chi^2$ for the fit.  Having such residues
one can then evaluate the counterterm contributions in terms of
\begin{eqnarray}
b^a&=&\left({\Lambda_\chi\over 2M_N}\right)\sum_Vb_V^a\nonumber\\
c^a&=&\sum_V\left({\Lambda_\chi\over m_V}\right)^2a_V^a
\end{eqnarray}
Jaffe then pointed out that one could evaluate the strange residues at the
$\omega,\phi$ poles in terms of the isoscalar electromagnetic residues
via\footnote{Note that Jaffe's analysis was framed purely in the
language of vector-dominance.  The chiral framework which we employ
here is taken from Ref.~\ref{ref:mi}.}
\begin{eqnarray}
{a_\omega^s\over a_\omega^{I=0}}&=&-{\sqrt{6}\sin\epsilon\over 
\sin(\epsilon+\theta_0)}\nonumber\\
{a_\phi^s\over a_\phi^{I=0}}&=&-{\sqrt{6}\cos\epsilon\over 
\cos(\epsilon+\theta_0)}\label{eq:ja1}
\end{eqnarray}
where $\epsilon\simeq 0.053\pm 0.005$ is the mixing angle between the 
physical $\omega,\phi$ and pure $\bar{s}s,(\bar{u}u+\bar{d}d)/\sqrt{2}$
states and is determined from the decay $\phi\rightarrow \pi\gamma$
\cite{hg,hsc}, 
while $\theta_0=\tan^{-1}1/\sqrt{2}$ is the ``magic'' angle of 
octet-singlet mixing that would yield such flavor-pure states.  
In order to obtain
the strange residues for the higher mass vector state, Jaffe employed
assumptions on the asymptotic dependence of the form factors -- that
$F_1$ vanish as $1/q^2$ and $F_2$ as $1/q^4$.  In this way he obtained
values
\begin{equation}
\kappa^s\simeq -0.3,\quad r_s^2 \simeq 0.2 \,\,{\rm fm}^2\label{eq:ja}
\end{equation}

While reasonable and seemingly somewhat model-independent, the difference
in sign between the strangeness charge radius obtained in this analysis
and that found in those described below should serve as a red flag.  In
addition, it should be 
emphasized that the values obtained in the Jaffe analysis are actually quite
sensitive to the mixing angle $\epsilon$ and to the assumed representation
of the form factors in terms of three poles, two of which are identified 
in terms of the physical $\omega$ and $\phi$ states.  It is particularly
sensitive to identification of the second resonance with the $\phi$ because 
of its very large strangeness content and its relatively large 
OZI-violating coupling to the nucleon.  In fact, the asymptotic 
dependence of the form factors which results from quark counting arguments
is even stronger than assumed
in this analysis and would require inclusion of additional dynamical 
structure.  One should also be concerned about the representation of the 
high energy
continuum structure in terms of a simple zero-width pole, since the
distribution of strength between strange and non-strange currents is
presumably an energy-dependent quantity.  One might hope, nevertheless, that 
low energy static properties should be relatively insensitive
to the couplings to high mass states so that perhaps Eq. \ref{eq:ja}
represents reasonable lowest order estimates.    In this regard, Forkel 
has shown that use of QCD asymptotics can reduce the size of 
the strangeness couplings by a 
factor of two to three, although the signs remain fixed.\cite{fork}

\subsection{Kaon loop models}

An alternate approach to the problem of strangeness matrix elements 
has been to abandon the requirement of a
consistent chiral expansion and to simply include a (hopefully reliable)
kaon loop contribution.  Of course, this must be done carefully since 
one can see from Eq. \ref{eq:lo} that a naive estimate of this type includes
divergences.  Thus some sort of cutoff procedure must be employed.
In fact, this is reasonable.  Indeed, recently Donoghue and Holstein (DH)
have pointed out that the feature that the baryon has an intrinsic size
strongly suggests the use of some sort of regularization which de-emphasizes
the effects from heavy meson (and therefore short distance) effects, since
such features must surely be suppressed by baryon structure.\cite{dh}
They advocated introduction of a dipole regulator
\begin{equation}
F(q^2)=\left({\Lambda^2\over \Lambda^2-q^2}\right)^2
\end{equation}    
into such chiral loop integrals involving baryons and showed that the
results of such a procedure retained the underlying chiral
structure while maintaining the phenomenological success
of lowest order chiral predictions, which are often obscured by large
nonanalytic contributions from chiral loops.  Of course, consistent
power counting is lost in such a scheme but the price may be worth paying.
The dipole regulator used by DH is similar to use of a form factor 
and therefore in many ways justifies the simple form-factor modified
kaon loop approaches which have appeared in the literature.  
One complication which must be faced when using a form factor or regulator
is the requirement of gauge invariance, which can easily be lost if one
is not careful.  In order
to retain the proper Ward-Takahashi identities it is necessary to include
some sort of contact or seagull contribution, whose form is ad hoc. 
For example, in the case of pseudovector meson-nucleon coupling
\begin{equation}
F(q^2)q^\lambda \phi \bar{u}\gamma_\lambda\gamma_5 u
\end{equation}  
one can insert a contact term such as\cite{mb,fn}
\beqna
&\ &\left\{iA_\mu(P^\mu+2k^\mu)\left({F((P-q)^2)-F(q^2)\over (P+q)^2-q^2}\right)
q^\lambda [\hat{Q},\phi]+iA^\lambda F((P+q)^2)[\hat{Q},\phi]\right\}\nonumber\\
&\cdot&\bar{u}\gamma_\lambda\gamma_5 u\label{eq:gi}
\eeqna 
whose form is, of necessity, model-dependent.\footnote{The Ward-Takahashi
identity constrains only the longitudinal component of the vertex.}
Another problem is the size of the cutoff or regulator term, $\Lambda$.
In the calculations of DH results were relatively insensitive to the precise
size of such a term as long as it was in the range 300 MeV $\leq\Lambda\leq$
600 MeV, but this is an unavoidable uncertainty since it represents short
distance effects which are omitted from the calculation.  A precise evaluation
of such effects would include additional physics and would be independent
of the regulator mass and to the extent that there is a strong 
dependence on $\Lambda$ such an approximate calculation should be judged to be uncertain.
(Thus it is worrisome that there is significant cutoff dependence in
the loop analysis performed, {\it e.g.}, in Ref. \ref{ref:mi}.)

In any case, various such calculations have been performed.  A common feature
of any such estimate is that the strangeness charge radius is predicted to
be negative since the kaon loop, which carries the $\bar{s}$, is further
from the center of mass than is the strange baryon, which carries the $s$
quark, and this is borne out in the calculations.  Note, however, as mentioned
above, that this is in disagreement with Jaffe's estimate and the simple
reasoning given here is based on static arguments and neglects recoil
effects.   One indication of the importance of such corrections can be
seen from the feature that if the cutoff mass is allowed to become very
large, corresponding to a pointlike kaon, the strangeness charge radius 
actually becomes {\it positive}, although for the $\Lambda\leq 1$ GeV
values used in the Bonn potential one finds $r_s^2<0$.  It should also
be noted that in general the magnitude of the
charge radius predicted in such models is found to be much smaller
than in the vector dominance approach.  There have been a number of 
such calculations, whose results are qualitatively similar and which
differ only in detail.  One such calculation is that of Ramsey-Musolf and 
Burkardt\cite{mb} 
who employed a simple $K\Lambda$ loop calculation using the phenomenological
meson-baryon form factors used by practitioners of the Bonn-J\"{u}lich
potential.\cite{bo}  (Effects from $K\Sigma$ 
and $\eta N$ loops were assumed to be much smaller, in accord with
SU(3) symmetry arguments, and were ignored.)  Seagull terms, as described
in Eq. \ref{eq:gi} were included in order to maintain the Ward-Takahashi
identities.  These authors find 
results consistent with Jaffe, {\it i.e.} a moderate negative value, 
for the strangeness moment but a smaller and 
{\it negative} value for the strangeness charge radius, as shown in 
Table~\ref{tab:modelResults}.

An alternative approach was taken by Pollock, Koepf, and Henley who 
employed a cloudy bag model (CBM) to estimate the strangeness 
moments.\cite{pkh}  The simple MIT bag model deals with the confinement issue
by incorporating a phenomenologically determined QCD vacuum energy into
a theory of noninteracting quarks.\cite{mit}  Bubbles (``bags'' of 
perturbative vacuum containing such quarks are then allowed to exist 
and are stabilized against collapse to the QCD vacuum phase by the pressure
exerted via the Heisenberg energy of such quark states.  A major failing of
the model, however, is that chiral symmetry is not maintained.  The CBM 
is a version of the MIT bag model which incorporates broken chiral symmetry 
by coupling to mesons at the surface of the bag.  In this way an
intrinsic cutoff parameter is set by the inverse size of the bag.  Results 
for many static properties of the nucleon has been calculated in such models
and results are generally successful.  In the case of the vector 
current strangeness matrix element by Pollock, Koepf, and Henley the
results are of the same order as found in the simple kaon loop picture,
though the strangeness magnetic moment is somewhat smaller.  Since the 
kaon loop calculation in Ref. \ref{ref:mb} utilizes chiral $K-\Lambda$
couplings and a cutoff parameter not dissimilar to the inverse bag
radius used in Ref. \ref{ref:pkh} it is perhaps not surprising that the
results are similar.

Cohen, Forkel, and Nielson proposed an approach which unites the
phenomenologically successful vector dominance ideas suggested by Jaffe
with the intuitively appealing meson cloud picture utilized
by Ramsey-Musolf and Burkhardt.  Emphasizing that simply adding the results of
previous calculations raises important questions involving double counting
they considered a ``hybrid'' model wherein ``intrinsic'' NN strange and
isoscalar matrix elements generated by kaon loops are ``renormalized''
via use of the current field identity together with phenomenological mixing
parameters to yield predictions for corresponding experimental 
quantities.\cite{cfn}  Their results are shown in Table~\ref{tab:modelResults}
and are
similar to the results obtained via kaon loop arguments.  Now in fact the
agreement in the case of the strangeness magnetic moment is required by the 
feature that this quantity is obtained at $q^2=0$ and consequently receives
no contribution from vector meson mixing -- it is given purely in terms of 
the ``intrinsic'' kaon loop diagrams.  The strangeness charge radius on the
other hand receives contributions both from the intrinsic component {\it and}
from the vector meson mixing.  The latter considerably enhances the 
very small negative value found in the kaon loop calculation by a factor
of two to three which makes the size much more reasonable from an 
experimental point of view.  As in the Jaffe analysis, however, this 
result is critically dependent on the size of the parameter $\epsilon$.
   
\subsection{Skyrme model}

One of the first pictures used in order to estimate 
strangeness effects was the Skyrme model.\cite{skr}  In the standard
SU(2) scheme the lowest order chiral 
Lagrangian, Eq. \ref {eq:abc} is augmented by a phenomenological 
four-derivative term
\begin{equation}
\Delta {\cal L}_4\sim{1\over 32 e^2}{\rm tr}[\partial_\mu UU^\dagger,
\partial_\nu U^\dagger]^2\label{eq:st}
\end{equation}
which stabilizes a classical solition solution
\begin{equation}
U_0=\exp[iF(r)\hat{x}\cdot\vec{\tau}]
\end{equation}
against collapse.  Allowing time-dependent quantum corrections around this
solution via $U=A^{-1}(t)UA(t)$ the Lagrangian reduces to
\begin{equation}
{\cal L}=-M+\lambda{\rm tr}[\partial_0A(t)\partial_0A^{-1}(t)]
\end{equation}
where, in terms of the dimensionless variable $\tilde{r}=2eF_\pi r$
\begin{eqnarray}
M&=&{8\pi F_\pi\over 3}\int_0^\infty\tilde{r}^2d\tilde{r}\left\{
{1\over 8}\left(F')^2+{2\sin F\over \tilde{r}^2}\right]
+{\sin^2F\over 2\tilde{r}^2}\left[
{\sin^2F\over \tilde{r}^2}+2(F')^2\right]\right\}\nonumber\\
\lambda&=&{\pi\over 3e^3F_\pi}\int_0^\infty\tilde{r}^2d\tilde{r}\left[
1+4\left((F')^2+{\sin^2F\over \tilde{r}^2}\right)\right]
\end{eqnarray}
This system may be solved using variational methods and what results is a
remarkably successful picture of nucleon structure.  Extension of the 
Skyrme approach to SU(3) in order to include strangeness effects is not 
straightforward and introduces additional model dependence.\cite{ext}  
The problem
was addressed in in Ref.~\ref{ref:psw} by appending terms involving nonmininal
derivative couplings into the Lagrangian in order to account for flavor
symmetry breaking.  After the usual quantization in the restricted space
of collective and radial excitations, the Hamiltonian is diagonalized while
treating the symmetry breaking terms exactly.  The results are shown in
Table~\ref{tab:modelResults} and reveal moderate negative values for both the 
strangeness charge
radius and magnetic moment.  

However, there is good reason to be concerned about the accuracy of such
predictions.  For one thing the extension to SU(3) involves a number of
ambiguities and the choice made by Park et al. in Ref. \ref{ref:psw} is
in that sense arbitrary.  Indeed the Skyrme model is generally 
acknowledged to be
justified in the large-$N_c$ limit, so that making reliable predictions
concerning subtle quark sea effects in the real $N_c=3$ world is a 
stretch.   Another indication of the approximate nature of any Skyrme
result is the fact that, as shown by Gasser and Leutwyler, the full four
derivative chiral Lagrangian consists of ten such terms and there are
even more contributions at six and higher derivatives which must be
included in order to fully represent the effective Lagrangian of QCD.
Thus any results arising from the use of the simple four-derivative 
stabilizing term Eq.~\ref{eq:st} can be approximate at best.  Finally, 
since the Skyrme practitioners evaluate the strangeness current
as the difference between the baryon number and hypercharge currents
\begin{equation}
J_\mu^s=J_\mu^B-J_\mu^Y
\end{equation}
any predictions for strangeness matrix elements are obtained by taking 
the (small) difference of two sizable and themselves uncertain quantities,
lending considerable doubt as to their reliability.  One indication of this 
uncertainty is that in a version of the model wherein vector mesons are
included, results become smaller by about a factor of two and the {\it sign} 
of the strangeness charge radius changes.\cite{pw}   

\subsection{Constituent quark approach}

Still another approach to the problem is to attempt to 
take the internal quark structure of the nucleon into account by 
representing it in terms of the usual complement of three {\it constituent} 
($U,D$) quarks, whose substructure consists partly of $\bar{s}s$ pairs.  
This is 
the calculational manifestation of the suggestions of Kaplan and
Manohar.\cite{KaMa}  The question is how to take this quark sea structure
into account.  A simple approach taken by Ramsey-Musolf and Ito is to assume
that the constitutent quarks are themselves coupled to mesons via a chiral 
quark model.\cite{mi}    
In this picture the strangeness content arises from the feature that the 
$U,D$ quarks can fluctuate into a kaon plus constituent $S$-quark.  
The axial coupling of the constituent quarks to mesons is determined by a 
constant $\lambda_A$ which is fit by demanding agreement of the chiral
quark model calculation with the experimental value of the axial
coupling in neutron beta decay -- $g_A\simeq 1.27$.  This procedure too is
fraught with model-dependent assumptions, however.  Indeed, since 
such loop calculations are themselves divergent, new counterterms 
\begin{equation}
\Delta{\cal L}={b_q^a\over 2\Lambda_\chi}\bar{\psi}\sigma_{\mu\nu}\hat{Q}
\psi F^{\mu\nu}-{c_q^a\over \Lambda_\chi^2}\bar{\psi}\gamma_\mu\hat{Q}
\psi\partial_\nu F^{\mu\nu},
\end{equation}
corresponding to a magnetic moment and charge radius for the constituent
quark, must be determined in terms of experimental quantities.  Since
the singlet channel quantities -- $a_q^0,b_q^0$ -- are unknown we have
no predictive power and are forced to assume something.  Using
arguments similar to those of DH, Ramsey-Musolf and Ito,
for example, inserted a cutoff $\Lambda_\chi$ into the loop integrations
in order to avoid infinities and assumed the validity of the simple
one kaon loop approximation, obtaining the results shown in 
Table~\ref{tab:modelResults}.
One unfortunate feature of this model, however, as noted by the
authors, is that in addition to 
the usual model-dependence one also has to be concerned about double
counting issues. Specifically, it is not clear whether $Q\bar{Q}$ bound states, 
which must be present in such a model, should be included in or separate
from the Goldstone bosons which are included as part of the chiral
couplings.

Another chiral quark model calculation, though somewhat more elaborate
since it includes both kaon as well as $K^*$ and $K^*-K$ loops, 
has been done by the Helsinki group.\cite{hel}  Inclusion of the vector loops 
somewhat stabilizes the calculation against senstivity to the size of
the regulator provided that a cutoff of order the chiral symmetry breaking 
scale is chosen.  The authors find a small negative ($\sim$ -0.05) 
value for the strange magnetic moment, with a substantial cancellation 
arising between the kaon and $K^*$ loop contributions.  In this work, 
besides the remaining cutoff dependence there is also an 
uncertainty introduced due to the loop diagram involving the $K-K^*-\gamma$ vertex. 
  
An alternative approach is to include strangeness
in the constituent quark structure via use of Nambu-Jona-Lasinio (NJL) model
methods.\cite{bjm}  The NJL model is an effective field theory involving
relativistic fermions interacting through four-point vertices subject to
the requirements of chiral symmetry.  In order to introduce a strangeness
component into the valence constituent quarks one requires some sort of 
flavor mixing interaction.  In the work of Refs.~\ref{ref:klv} 
and~\ref{ref:vlk} this
was accomplished within the mean field (Hartree-Fock) approximation to
the NJL model, wherein such terms arise from six-quark interactions 
involved in the determinant.  Such terms are necessary in order to provide the
anomalous breaking of $U_A(1)$ symmetry implied by 't Hooft's instanton
methods\cite{to} and involve, in general, constants which must be fit 
empirically in terms of known singlet-octet and $\rho-\omega$ mixing.
There exists considerable model dependence in this process but the $\bar{s}s$
component which is allowed in the $U,D$ quarks is generally small.
A specific estimate of the strangeness charge radius by Forkel et al.
yields a small positive value of 0.017 fm$^2$.\cite{fn}  
A related alternative approach is to bosonize the
NJL model, yielding an effective Lagrangian containing a topological 
soliton, which may then be solved via Skyrme methods.\cite{wi}  In this case
also there exists considerable model dependence as can be
seen from the range of numbers which are allowed by reasonable assumptions
({\it cf.} Table~\ref{tab:modelResults}).

\subsection{Dispersion relations}

Another approach to the subject of strangeness content is through dispersion
relations.\cite{mhd}  In many ways this technique is like that of 
effective field theory,
since in both cases one takes a method based on general principles -- chiral
symmetry in the case of chiral perburbation theory and causality and
analyticity in the case of
dispersion relations in order to interrelate experimental quantities, although
an important difference is that the dispersion relations do not involve a
systematically controlled approximation.  In
the case of the strangeness content, for example, one has the relations
\begin{eqnarray}
r_s^2&=&{6\over \pi}\int_{t_0}^\infty dt{{\rm Im}F_1^s(t)\over t^2}\nonumber\\
\kappa^s&=&{1\over \pi}\int_{t_0}^\infty dt {{\rm Im}F_2(t)\over t}
\end{eqnarray}
Here the possible intermediate states which contribute to the spectral
function Im$F_1^s(t)$ include $3\pi,K\bar{K},N\bar{N}$, etc.  
It is suggestive to focus on the $K\bar{K}$ in particular since this
is the lightest state which has no OZI rule suppression.  An important
complication is the feature that the required spectral density
utilizes contributions from $N\bar{N}\rightarrow K\bar{K}\rightarrow 
V_\mu$ in the region $4m_K^2,t,\infty$, but direct data exist only 
when $t>4M-N^2$.  The solution to this problem is to invoke crossing 
symmetry and backward dispersion relations in order to produce 
the needed helicity amplitudes for $N\bar{N}\rightarrow K\bar{K}$ in
terms of measured $KN$ scattering data.  This program was carried out
by Ramsey-Musolf and Hammer\cite{rmh} who expressed the required
spectral densities via
\begin{eqnarray}
{\rm Im}F_1^s(t)&=&\left({M_N\sqrt{{t\over 4}-m_K^2}\over t-4M_N^2}\right)
\left[\sqrt{1\over 2}{E\over M_N}b_1^{1/2,-1/2}-
b_1^{1/2,1/2}\right]F_K^s(t)^*
\nonumber\\
{\rm Im}F_2^s(t)&=&-\left({M_N\sqrt{{t\over 4}-m_K^2}\over t-4M_N^2}\right)
\left[\sqrt{1\over 2}{M_N\over E}b_1^{1/2,-1/2}-
b_1^{1/2,1/2}\right]F_K^s(t)^*
\end{eqnarray}
where $b^{1/2,\pm 1/2}$ are the helicity amplitudes for the process $K\bar{K}
\rightarrow N\bar{N}$ obtained from analytic continuation of 
$KN$ scattering data, while $F_K^s(t)$ is the kaon strangeness form factor, 
which is defined through the matrix element
\begin{equation}
<0|\bar{s}\gamma_\mu s|K(p_1)\bar{K}(p_2)>\equiv F_K^s(t)(p_1-p_2)_\mu
\end{equation}
for which no data exists.  Thus one must of necessity make some model
dependent assumptions in order to perform the requisite dispersive 
integration.  Ramsey-Musolf and Hammer addressed this problem by
looking at the related isoscalar electromagnetic form factor where one
has data on the {\it vector 
current} $K\bar{K}$ coupling from electron
scattering.  It is known in this case that the isoscalar magnetic 
from factor in particular can be well described in terms of opposite
sign contributions from $\omega$ and $\phi$ states plus small
corrections from higher mass states.  (The same is true, but to a
lesser extent, for the $q^2$ dependence of the charge form factor.)
The authors noted that the residues in such a vector-meson pole model 
fit to the isoscalar electromagnetic form factors could be carried
over to the desired strangeness problem using the Jaffe rotation, 
Eq.~\ref{eq:ja1}.  Since
\begin{equation}
{a_\omega^s\over a_\omega^{I=0}}\sim -0.2;\quad {a_\phi^s\over
a_\phi^{I=0}}\simeq -3
\end{equation}
we see that the $\omega$, which is associated with the $3\pi$
intermediate state, can essentially be neglected, while the $\phi$ 
plays a dominant role.  In this way these authors argued convincingly
that the low energy component of the dispersion integral is fairly
reliably estimated, leading to a small negative value for $\kappa_s$
consistent with original Jaffe estimate.  In the case of the
strangeness radius, there is somewhat more model dependence.  There
is also some uncertainty with respect to higher mass contributions.
With these caveats the results are listed in Table~\ref{tab:modelResults}.
Note that the 
strange magnetic moment is roughly consistent with the chiral
analyses, while the predicted charge radius is fairly large and 
positive, $r_s^2\sim$ 0.42 fm$^2$.

\begin{table}
\begin{center}
\begin{tabular}{c|c|c}
Model& $r^2_s$ (fm$^2$)& $\kappa_s$ \\
\hline\\
Vec. Dom.\cite{ja} & 0.2 & -0.3 \\
$K\Lambda$ loop\cite{mb} & -0.007 & -0.35 \\
CBM\cite{pkh}& -0.011 & -0.10 \\
Hybrid\cite{cfn}& -0.025 & -0.3 \\
chiral quark\cite{mi}&-0.035& -0.09\\
Skyrme\cite{psw}& -0.10--\,-0.15& -0.13--\,-0.57\\
Skyrme+VDom. & +0.05&-0.05\\
NJL-sol.\cite{wi}&-0.25--\,-0.15&-0.05--\,+0.25\\
Dis. Rel.\cite{mhd}& 0.42&-0.28
\end{tabular}
 \caption{Calculated values for the strangeness moment and charge radius
in various models.}
\label{tab:modelResults}
\end{center}
\end{table}

\subsection{Summary of theoretical models}

We have seen above that a variety of theoretical approaches have been applied
to the problem of calculating matrix elements of $\bar{s}{\cal O}s$
operators.  This survey is not meant to be exhaustive -- indeed there are
additional calculational approaches which we have not examined.  However,
it should be clear from our discussion that any such evaluation
must of necessity involve significant model dependence.  Indeed one 
indication of this 
fact is that the vector dominance analysis of Jaffe, predicts a 
{\it positive} strangeness radius, while most of the others come down on
the negative side.  Other
warning signs are present, too.  In each estimation there exist built-in
and inescapable assumptions concerning values for cutoff parameters, 
unconstrained coupling constants, etc. as well as the validity of
the simple one loop approach.  Although the dispersive evaluation 
seems somehow more secure, both in it as well as in the Feynman diagrammatic 
approaches, the authors strongly emphasize the model 
dependent features which go into their calculations, so that uncertainties 
are easily at the $\sim$100\% level.  This is simply the state of such
hadronic calculations at this time.  In spite of all the differences and
warning signs, however, most of the calculations appear to lead to similar
conclusions -- a moderate negative strangeness moment, $\kappa^s\sim -0.3$
and a small negative strangeness radius, $r_s^2\sim -0.010$ fm$^2$.

Lest we become too confident, however, it should be noted that the negative
signs for both the strangeness radius and magnetic moment favored by 
most of the models is {\it opposite} in sign to the central numbers 
measured by SAMPLE and HAPPEX (see Section 6).  An additional warning sign comes from 
the realization that the basis of the above loop calculations is that

\begin{itemize}
\item [i)] despite large coupling, rescattering (multi-loop) effects are
suppressed.

\item [ii)] the lightest hyperon-strange meson intermediate states 
generate the 
dominant contribution to the strangeness matrix elements;
\end{itemize}

Both assertions have been challenged by recent calculations.
Questions about the validity of the former are raised
by the dispersive evaluation of Ramsey-Musolf, Hammer, and 
Drechsel where rescattering
effects are found to be important even at relatively low energies and 
in order to build up resonance strength in the $\phi$ region.  In the case 
of the latter, a recent calculation by Geiger and Isgur, demonstrated 
in a simple model calculation that $\bar{s}s$ pair effects result from 
delicate cancellations
between much larger contributions from a significant number of virtual 
meson-baryon intermediate states rather than being dominated by only the 
lowest-lying, as assumed in most of the above analyses.\cite{gi}  This worry 
is consistent with, and amplified by, a calculation by Barz et al.,\cite{ba} which 
argues that $K^*$-loop effects, whether calculated via dispersive or Feynman 
diagram methods, are certainly comparable to or in some cases even larger 
than the corresponding kaon loop effects.\footnote{This may not be as
problematic as it originally seemed since recent work by Forkel, Navarra, and
Nielsen using the softer $K^*$ form factors favored by recent Bonn-J\"{u}lich
potential practitioners has substantially weakened these $K^*$-loop
effects.\cite{fnn}}  The problem may be that present calculations,
which rely 
generally on the assumption of dominance by a few light intermediate 
states and of neglecting rescattering, are insufficient.   If this is 
borne out by future work, the task of theoretically estimating such 
quantities and of making contact between experimental numbers and theoretical 
input for observables involving non-valence sea effects will indeed become 
even more daunting than it is already.  This, of course, is simply an 
indication of the difficulties of dealing with subtle issues of 
hadronic physics.  Its solution 
awaits reliable unquenched lattice calculations and/or improved 
hadronic models/methods and remains one of the important challenges
for study of the strong interaction as we enter
the twenty-first century.  

\section{Experimental Methods} 
\noindent
Since, as described above, there is no convincing way to calculate the
strange quark contributions, it must be the task of experiment to provide
the information. 
The strange quark matrix elements described in Section 3 are
determined from comparison of the electromagnetic and weak currents as
discussed in Section 2.1.  The neutral weak currents are
measured in parity-violating asymmetry measurements as discussed in
Section 2.3.  The experimental requirements imposed by the small
asymmetries in these measurements center on three areas: statistical
precision, systematic accuracy and kinematic selection.  The present
experiments take advantage of the pioneering parity-violation
measurements at SLAC\cite{Prescott}, MIT-Bates\cite{C12} and
Mainz.\cite{Be9} There are a number of different realizations in the
present experimental program which will be summarized briefly below.

Because the asymmetries range from $\sim 10^{-5} - 10^{-4}$ for the $Q^2$
range of current experiments ($0.1 < Q^2 <1$ GeV$^2$), there is a
strong requirement for high counting rates.  All experiments use a
liquid hydrogen or deuterium target and an intense electron beam.
With target lengths of 10 - 40 cm and beam currents of 40 - 100 $\mu
A$, the typical luminosity in these experiments is ${\cal L} \ge
10^{38}$ cm$^{-2}$ s$^{-1}$.  In principle, the largest detector solid angle
possible is also desirable.  This may be done directly as in the
SAMPLE experiment, where the acceptance at backward scattering angles is
$\Delta\Omega \sim 1.5$ sr, or a large fraction of the azimuthal
acceptance may be covered at very small scattering angles as is the
case with the HAPPEX experiment where the solid angle acceptance is
about 11 msr.  It should also be noted that the standard figure of merit for these
asymmetry measurements
\beqn
F = P_{beam}^2 A^2_{PV} \sigma \propto \frac{1}{T_{expt}}
\eeqn
is essentially independent of momentum transfer (as $A_{PV}\propto
Q^2$ and $\sigma \propto Q^{-4}$) at low momentum transfers where the
form factors are of order 1.  The experimental techniques are
therefore roughly independent of momentum transfer.  In order to
measure an asymmetry of $10^{-5}$ with an uncertainty of a few
percent, typically 45 days of beam time are required.  This time
depends critically on the beam polarization of course: the uncertainty
achieved is directly proportional to $1/P_{beam}$.

With statistical precision approaching $\Delta A_{stat} \sim 10^{-7}$,
the clear challenge is to achieve comparable accuracy in the
systematic corrections that must be applied to the measurements.  In
general, the measured asymmetry must be corrected.  To first order
\beqn
A_{phys} = A_{meas} - \sum_{i} \frac{1}{2}\frac {\partial ln(Y)}
{\partial x_i} \Delta x_i
\eeqn
where $Y$ is the measured yield and the $x_i$ are the beam parameters
such as position, angle, energy, size, etc.  In particular, it is the
helicity-correlated differences in these beam parameters, $\Delta
x_i$, that are necessary to correct the measured asymmetry.  In
practice, for electron scattering experiments, both the
helicity-correlated beam parameter differences and the yield
derivatives are measured continuously during the experiment.  In order
to measure the derivatives two methods may be employed.  The natural
variation of the beam parameters during a given measurement interval
may be used to construct the derivatives to correct the data for that
interval, or the beam parameters may be deliberately varied over a
wider range and the resulting derivatives utilized.

Of course if the helicity-correlated beam parameter variations vanish,
then no corrections would be necessary.  Reducing the $\Delta x_i$ has
required substantial effort, but the results have been encouraging.
In fact, generating polarized electron beams that essentially differ
only in their helicity is possible because of the exquisite control
possible in modern laser beam lines.  Polarized electron sources
produce beams using the photoelectric effect with circularly polarized laser
light.  Electrons with positive and negative helicities are generated
from either bulk (maximum polarization 50\%) or strained (maximum
practical polarization up to $\sim 90\%$ at reduced currents) GaAs using
right- and left-circularly polarized light, respectively.  The light
polarization is reversed using an electro-optic $\lambda/4$ plate
whose axes can be rotated rapidly (Pockels cell).  Feedback loops are
typically used to control the intensity of the right- and left-handed
light beams as well as their positions on the photocathode.  Great
care must be taken, especially with the strained crystals, to reduce
the linearly polarized component of the light (which generally changes
magnitude and/or direction when the circular polarization is
reversed) because the strained crystal acts as an optical analyzer.  
Typical circular polarizations are 99.95\% and the goal
for the most effective operation with strained crystals is 99.999\%.\cite{sinclair}
Using present techniques, helicity-correlated beam position differences
at the target are typically reduced to $~10$ nm or less and relative
helicity-correlated energy differences to $10^{-8}$.  It should also 
be noted that in contrast to proton parity-violation experiments where
the beams are non-relativistic, the effects of transverse polarization
are small in the electron case, since they are generally reduced by the relativistic
$\gamma$ factor.
 
Because of the large acceptances required to measure these small
asymmetries, some attention must be paid to selection of kinematics.
The natural variation of the electron scattering cross section is
favorable in this case.  The general goal is to measure a particular
piece of the asymmetry at a particular momentum transfer.  At backward
electron scattering angles, where the cross section is small, the
momentum transfer varies slowly with angle allowing a large solid
angle acceptance to be used.  At forward scattering angles, the
momentum transfer varies rapidly with angle, necessitating a
relatively smaller acceptance.  However, the forward cross sections
are much larger and offset the acceptance.  A complete separation of
the three neutral weak form factors, $G_E^Z$, $G_M^Z$, and $G_A^Z$
requires either measurements at three angles or, more practically, 
measurements at two angles and a
measurement of quasi-elastic scattering from deuterium where the
combination of asymmetries from the proton and neutron enhance the
axial form factor over the strange quark (``unknown'') parts of the
vector form factors.

\section{Experimental Results}		
\noindent

Following three pioneering parity-violating electron scattering
experiments,\cite{Prescott,C12,Be9} a program has developed to investigate primarily the
elastic neutral weak form factors of the nucleon.  Two experiments
have published results: SAMPLE and HAPPEX.  These experiments will
continue and be supplemented by several new experiments (including PVA4 and G0)
which will be discussed briefly in the following section.

\subsection{SAMPLE experiment}

\begin{figure}[htbp]
\vspace*{13pt}
\vspace*{2.75truein}		
\vspace*{13pt}
\includegraphics{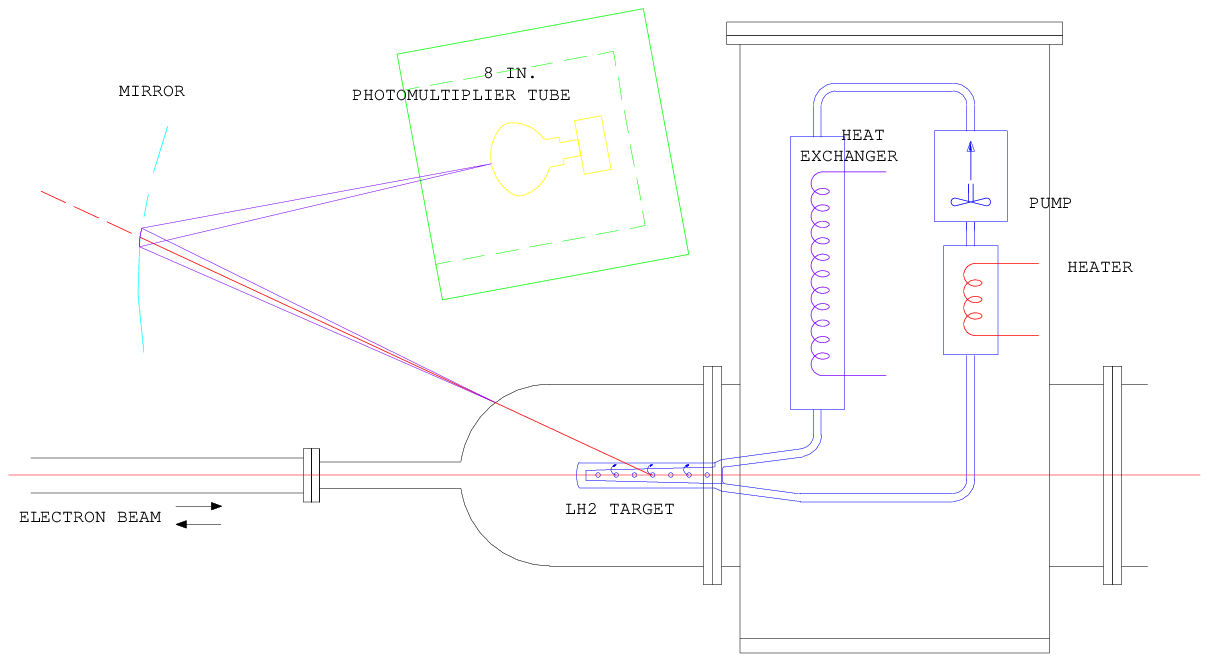}
\fcaption{Schematic representation of the SAMPLE experiment.  The
detector consists of 10 mirror-phototube pairs to collect the Cerenkov
light emitted in air by the scattered electrons.}
\label{fig:sampleSchematic}
\end{figure}

In the SAMPLE experiment, 
elastic scattering from the proton\cite{mueller,spayde} and
quasi-elastic scattering from the deuteron\cite{hasty} are measured at backward
angles as shown schematically in Fig.~\ref{fig:sampleSchematic}.  
The average momentum transfer of the measurement is 0.1
GeV$^2$, with scattering angles ranging between about 130$^\circ$ and
170$^\circ$.  The polarized beam from the MIT-Bates accelerator
($P_{beam} \sim 35\%$ using a bulk GaAs photocathode) with an
intensity of 40 $\mu$A is incident on a 40 cm liquid hydrogen target.
The scattered electrons are measured using a 10 segment air Cerenkov detector
(threshold energy of 21 MeV).  Because the instantaneous counting rate
in a given detector segment is about 100 MHz, the signals are integrated.

The measured proton asymmetry is
\beqn
A_p^{exp}(Q^2=0.1 \hbox{GeV}^2 \hbox{, }\theta_{av}=146.2^\circ) = -4.92
\pm 0.61 \pm 0.73~\hbox{ppm}
\eeqn
where the first uncertainty is statistical and the second
systematic. This is to be compared to a ``standard'' asymmetry of
-7.2~ppm where it is assumed that the strange quark contribution to
the vector currents is zero and that the corrections to the axial
current are as calculated in Ref.~\ref{ref:mrm}.  Showing the dependence on the
strange quark and axial contributions explicitly, the theoretical asymmetry is
\beqn
A_p^{th}(Q^2=0.1 \hbox{GeV}^2 \hbox{, }\theta_{av}=146.2^\circ) = (-5.61 +
1.37 G_A^e(T=1) + 3.49 G_M^s)\,\,{\rm ppm}
\eeqn

The systematic uncertainty in this measurement is dominated by
contributions other than those associated with false asymmetries.  The
largest contributions are from background subtractions.
Background subtractions for this experiment involve both light from
scintillation rather than Cerenkov radiation as well as yield not
associated with light (measured with a shutter in front of the each
phototube).  In particular for the proton measurement, the ``shutters
closed'' and ``scintillation'' asymmetries contributed essentially all of
the 15\% systematic uncertainty.  In contrast, the individual
contributions from the beam parameter corrections (``false
asymmetries'') and the beam polarization were about 5\%.

A separate measurement was performed with the same apparatus but with a
deuterium target.  Because of the large energy acceptance of the
detector both elastic and quasi-elastic scattering from the deuteron
were measured.  The elastic scattering and threshold electrodisintegration contributions
(based on the appropriate fractions of the yield) were estimated to 
change the measured asymmetry by only about 1\%.  The measured asymmetry was
\beqn
A_d^{exp}(Q^2=0.1 \hbox{GeV}^2 \hbox{, }\theta_{av}=146.2^\circ) = -6.79
\pm 0.64 \pm 0.51~\hbox{ppm}
\eeqn
In this case the expected asymmetry is -8.8 ppm again assuming zero
strange quark contribution and the ``standard'' axial corrections of
Ref.~\ref{ref:mrm}.  Again showing the theoretical asymmetry explicitly in terms of the strange
quark and axial contributions
\beqn
A_d^{th}(Q^2=0.1 \hbox{GeV}^2 \hbox{, }\theta_{av}=146.2^\circ) = 
(-7.60 + 1.93 G_A^e(T=1) + 0.88 G_M^s)\,\,{\rm ppm}
\eeqn
It can be seen here that whereas the sensitivity of the deuterium
measurement to the axial current is similar to that of the proton, the
contribution from the strange quarks is significantly smaller (essentially by
the ratio $(\mu_p+\mu_n)/\mu_p$).  The systematic uncertainties in
this measurement are essentially the same as those for hydrogen with a
smaller overall ``shutters closed'' uncertainty.
 
\begin{figure}[htbp]
\vspace*{13pt}
\vspace*{3.0truein}		
\vspace*{13pt}
\includegraphics{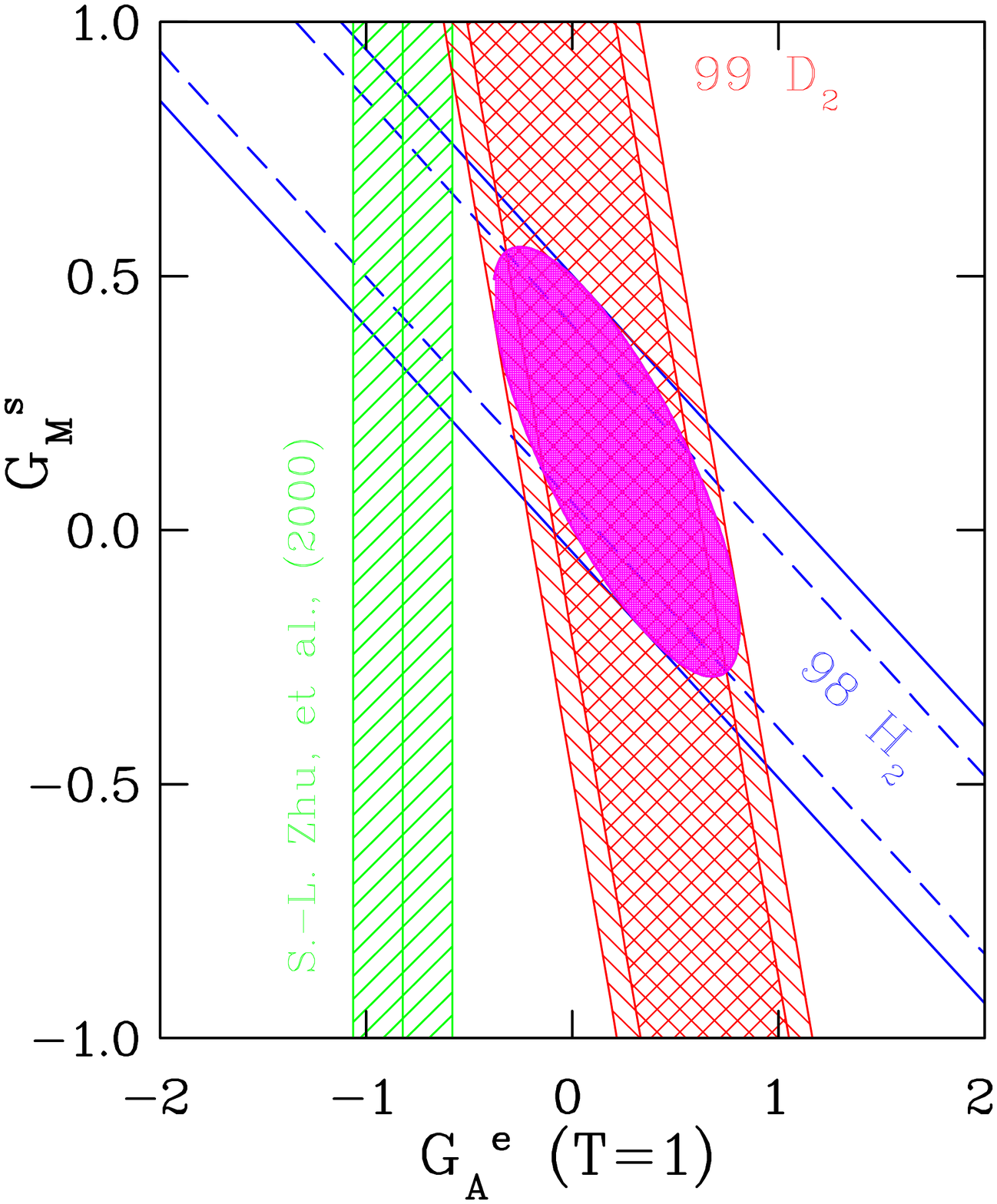}
\fcaption{Results from the SAMPLE measurements of parity-violating electron 
scattering on hydrogen\cite{spayde} (elastic) and deuterium\cite{hasty} 
(quasi-elastic).  The 
contribution of strange quarks to the proton magnetic moment at $Q^2=0.1$ 
GeV$^2$, $G_M^s$ is plotted vs. the effective isovector axial current seen by 
the electron, $G_A^e(T=1)$.  
The vertical band is the theory of Ref.~\ref{ref:mrm}.}
\label{fig:sampleResult}
\end{figure}

The results of these measurements are shown in Figure
\ref{fig:sampleResult} plotted as a function of the strange quark
contribution to the magnetic form factor, $G_M^s$ and the isovector
axial current seen by the electron, $G_A^e(T=1)$.  It can be seen that
the result is rather far from the expectation of $G_M^s \sim
-0.3$ and $G_A^e(T=1)=-0.71 \pm 0.20$.\cite{mrm}  Instead the experimental results for
these quantities are
\beqn
G_M^s(Q^2=0.1 \hbox{GeV}^2) = 0.18 \pm 0.30 \pm 0.30
\eeqn
and
\beqn
G_A^e(T=1, Q^2=0.1 \hbox{GeV}^2) = +0.27 \pm 0.46 \pm 0.38
\eeqn

These results may also be cast in terms of the general static electron-quark
axial couplings $C_{2u}$ and $C_{2d}$\cite{PDG99}, which are defined
as coefficients in the axial part of the neutral current {\it eN} lagrangian
\beqn
{\cal L}^Z_A = \frac{G_F}{2}\sum_i C_{2i}\bar e \gamma_\mu e
\bar q_i \gamma^\mu \gamma^5 q_i
\eeqn
where $i$ runs over the quark flavors. As is the case
with $G_A^e$, the combination of the proton and deuterium measurements
is sensitive to the isovector combination $C_{2u} - C_{2d}$
\beqna
C_{2u} - C_{2d} &=& -(1 - 4 \sin^2\theta_W)\frac{(1+Q^2/M_A^2)^2}{g_A} 
G_A^e(T=1)\\ 
&=& +0.015 \pm 0.032 \pm 0.027
\eeqna
As would be expected from this equation, the tree level
contribution to $C_{2u} - C_{2d}$ is $-(1 - 4 \sin^2\theta_W)=-0.075$.
Adding in the corrections of Ref.~\ref{ref:mrm} changes the
expectation to $-0.058\pm0.02$, still about 1.5$\sigma$ from data
(cf. the roughly 1.5$\sigma$ deviation of $G_A^e$ from the
calculation in Fig.~\ref{fig:sampleResult}). 

A new deuterium measurement is planned\cite{tito} at lower momentum
transfer to improve the determination of the
axial current and check these results.  By measuring with an incident energy of 120 MeV rather
than 200 MeV, $Q^2$ and hence the asymmetry are reduced by a factor of
nearly three.  However, the cross section will increase significantly
relative to the earlier measurement while the background is expected
to stay more or less fixed.  Because the momentum transfers are so low
in both the original and new measurements, it is reasonable to expect
the form factors will change in a smooth and predictable manner in
this range of momentum transfer.  Therefore, in addition to providing
an additional measurement of essentially the same physics, some of the
critical experimental factors will change thus providing an important
cross check.

\subsection{The HAPPEX experiment}

The HAPPEX experiment\cite{hap,HAPPEX2} utilized the two
spectrometers in Hall A at Jefferson Lab to measure parity-violation
in elastic electron scattering at very forward angles.  In this case,
the relatively small solid angle of each spectrometer, $\Delta\Omega =
5.5$ msr, is compensated by the very large (0.7 $\mu$b/sr ) 
cross section at forward angles
($\theta=12.3^\circ$) yielding, with a 15 cm long liquid hydrogen target,
a rate of roughly 1 MHz 
in each spectrometer.  The
scattered electrons were detected by integrating the output of a
simple lead-scintillator calorimeter.  This calorimeter was shaped to
accept only the elastic electrons, which are physically well separated
from the inelastic electrons in the focal plane of the spectrometer.  The
experiment was performed in two stages.  The first used a 100
$\mu$A beam with 39\% polarization produced from a bulk Ga-As crystal.
In the second, a strained Ga-As crystal was used, resulting in a beam
polarization of about 70\% and a current of 35 $\mu$A, slightly
improving the overall figure of merit (P$^2$I).

\begin{figure}[htbp]
\vspace*{13pt}
\vspace*{2.81truein}		
\includegraphics{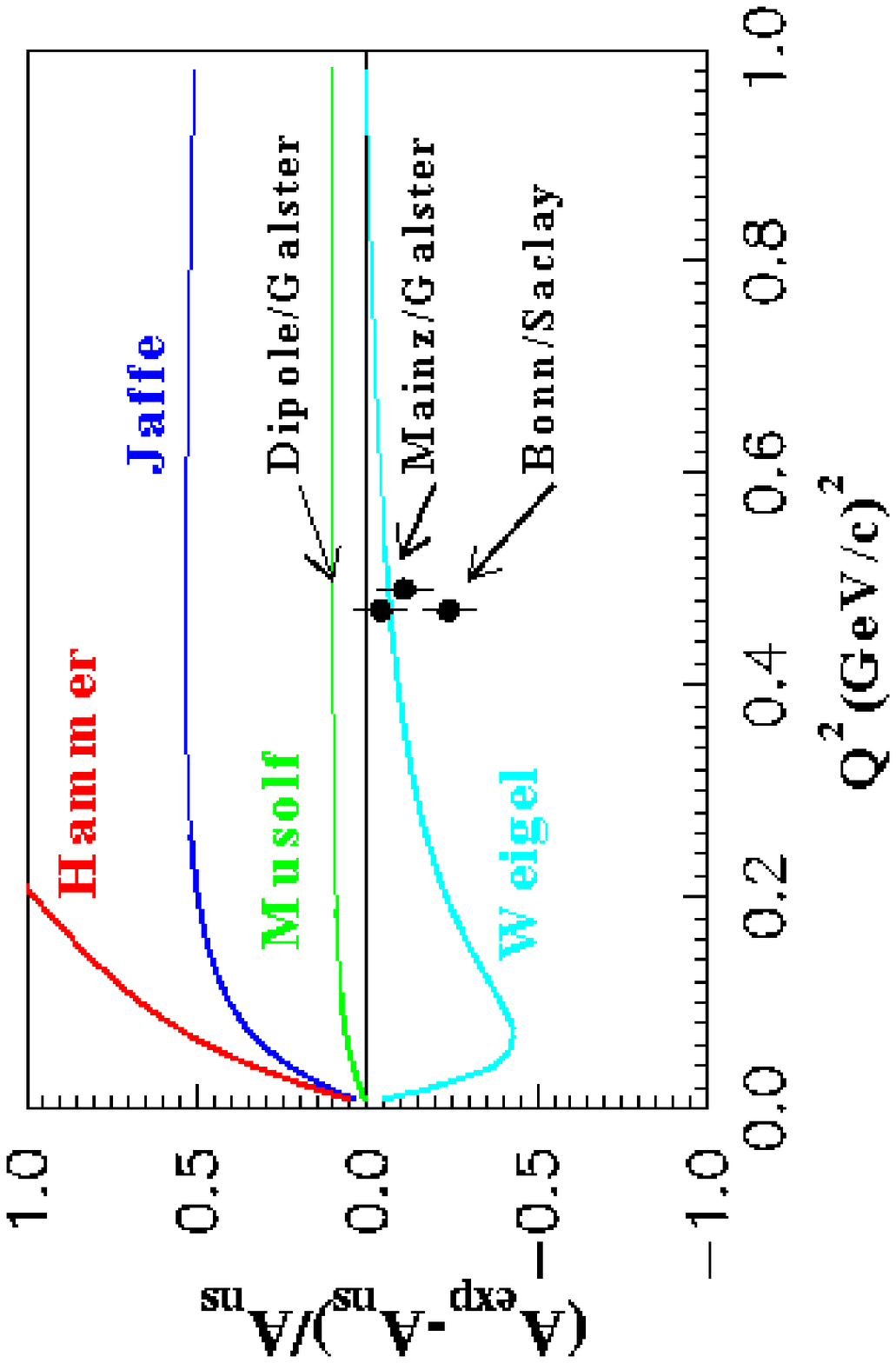}
\vspace*{13pt}
\fcaption{Results from the HAPPEX measurements of parity-violating electron 
scattering on hydrogen.  The three points on the plot correspond to
the neutron electromagnetic form factors of upper: $G_E^n$ from
Galster\cite{galster} and dipole $G_M^n$; middle $G_E^n$ from
Galster and $G_M^n$ from a Mainz
measurement\cite{mainzGMn}; and lower: $G_E^n$ from
a Saclay measurement\cite{platchkov} and $G_M^n$ from a Bonn
measurement\cite{Bonn}.  The calculations are from Hammer et
al.\cite{hammer}, Jaffe\cite{ja}, Musolf and Ito\cite{mi}, and Weigel,
et al.\cite{weigel}}
\label{fig:happexResult}
\end{figure}

The measured asymmetry, including the results from both phases of the
experiment is
\beqn
A_p(Q^2=0.477 \hbox{GeV}^2 \hbox{, }\theta_{av}=12.3^\circ) = -14.60
\pm 0.94 \pm 0.54~\hbox{ppm}
\eeqn
where again the first uncertainty is statistical and the second
systematic.  The largest sources of systematic uncertainty are
measurement of the beam polarization (3.2\% of its value) and
determination of $Q^2$ accruing from uncertainty in measurement of the
scattering (spectrometer) angle (contributing a 1.8\% uncertainty to
$A_p$).  There is a significant uncertainty in the result owing from
uncertainty in the neutron form factors as is shown in
Fig.~\ref{fig:happexResult} and discussed further below.  It should be
noted that Fig.~\ref{fig:happexResult} suggests that the magnitude of
the measured asymmetry is less than that with no strange quarks, in
accord with the SAMPLE result.

As is the case for the SAMPLE experiment, the systematic uncertainties
due to false asymmetries are very small---negligible, in fact, in this
experiment!  In principle, the helicity-correlated beam property
differences are more pronounced with the strained crystal because it
acts as an analyzer of linearly polarized light (always present at
some small level in the nominally circularly polarized laser 
beam---cf. Section 4).  However, intensity asymmetries were nulled with a
feedback system and position differences were reduced to an acceptable
level by tuning the optics of the laser line and by maximizing the
``transverse demagnification'' of the accelerator.  In the HAPPEX
experiment, sensitivity to these beam parameter variations was
measured by deliberately modulating the beam position and energy.  The
total correction for beam induced false asymmetries amounted to only
$0.02 \pm 0.02$ ppm or about 0.1\% of $A_p$.

Because the HAPPEX asymmetry was measured at a forward angle, it is in
principle sensitive to three unmeasured form factors - $G_E^s$,
$G_M^s$ and $G_A^e$.  The axial contribution is relatively small for
forward angles (becoming zero at $0^\circ$) and amounts to $-0.56 \pm
0.23$ ppm out of the total of -14.6 ppm, assuming the calculated
value\cite{mrm} for $G_A^e(T=1)$ rather than that measured in the
SAMPLE experiment.  The other form factors enter in the combination
$G_E^s + 0.392 G_M^s$ for these kinematics.  The value of this
combination, normalized to the most accurately measured proton form
factor, $G_M^p/\mu_p$, is 
\beqn 
\frac{G_E^s + 0.392 G_M^s}{G_M^p/\mu_p}
= 0.091 \pm 0.054 \pm 0.039 
\eeqn 
where the first uncertainty is a
combination of the statistical and systematic uncertainties in the
asymmetry combined in quadrature with the uncertainty in the axial
contribution, and the second is due to the uncertainty in the other
electromagnetic form factors.\cite{HAPPEX2} The results are
particularly sensitive to the neutron magnetic form factor as can be
seen in Fig.~\ref{fig:happexResult}; using the
results from a different recent $G_M^n$ measurement\cite{Bonn} yields
\beqn \frac{G_E^s + 0.392 G_M^s}{G_M^p/\mu_p} = 0.146 \pm 0.054 \pm
0.047 \eeqn Fortunately a number of experiments are planned to reduce
the uncertainty in $G_M^n$.  

\subsection {Future experiments}

The HAPPEX group is also approved to make
a measurement\cite{HAPPEXII} at a momentum transfer of $Q^2=0.1$ GeV$^2$ utilizing
new septum magnets placed in front of the existing spectrometers.
These septa will allow measurements at more forward angles (roughly
$6^\circ$ scattering angle) in order to increase the cross section at
low momentum transfers and
hence the overall figure of merit.  Other recently approved or re-approved
parity-violation measurements at JLab include one to determine the
neutron radius of the Pb nucleus\cite{Michaels} and a second to
measure the asymmetry in scattering from He at low momentum
transfer\cite{Armstrong} to measure the proton strangeness radius together 
with a He measurement at high momentum transfer\cite{Beise} 
($Q^2=0.6$ GeV$^2$) where early predictions showed a large value of $G_E^s$.
We note that because He is a $0^+$, $T=0$ nucleus, there are neither 
contributions from $G_M$ nor from $G_A$, making it particulary advantageous 
for measurements of $G_E^s$.

Two new parity-violation experiments are being mounted with dedicated 
apparatus to address the
questions of the weak neutral current in the nucleon.  The PVA4
experiment,\cite{pva4} underway at the MAMI accelerator in Mainz will
measure both forward and backward asymmetries using an array of
PbF$_2$ calorimeter crystals.  The G0 experiment,\cite{g0} to be performed at
JLab, will also
measure at forward and backward angles to 
separate the contributions of the charge, magnetic and axial terms
over the full range of momentum transfers from about $Q^2=0.1$ to $Q^2=1.0$
GeV$^2$.

\begin{figure}[htbp]
\vspace*{13pt}
\vspace*{3.0truein}		
\includegraphics{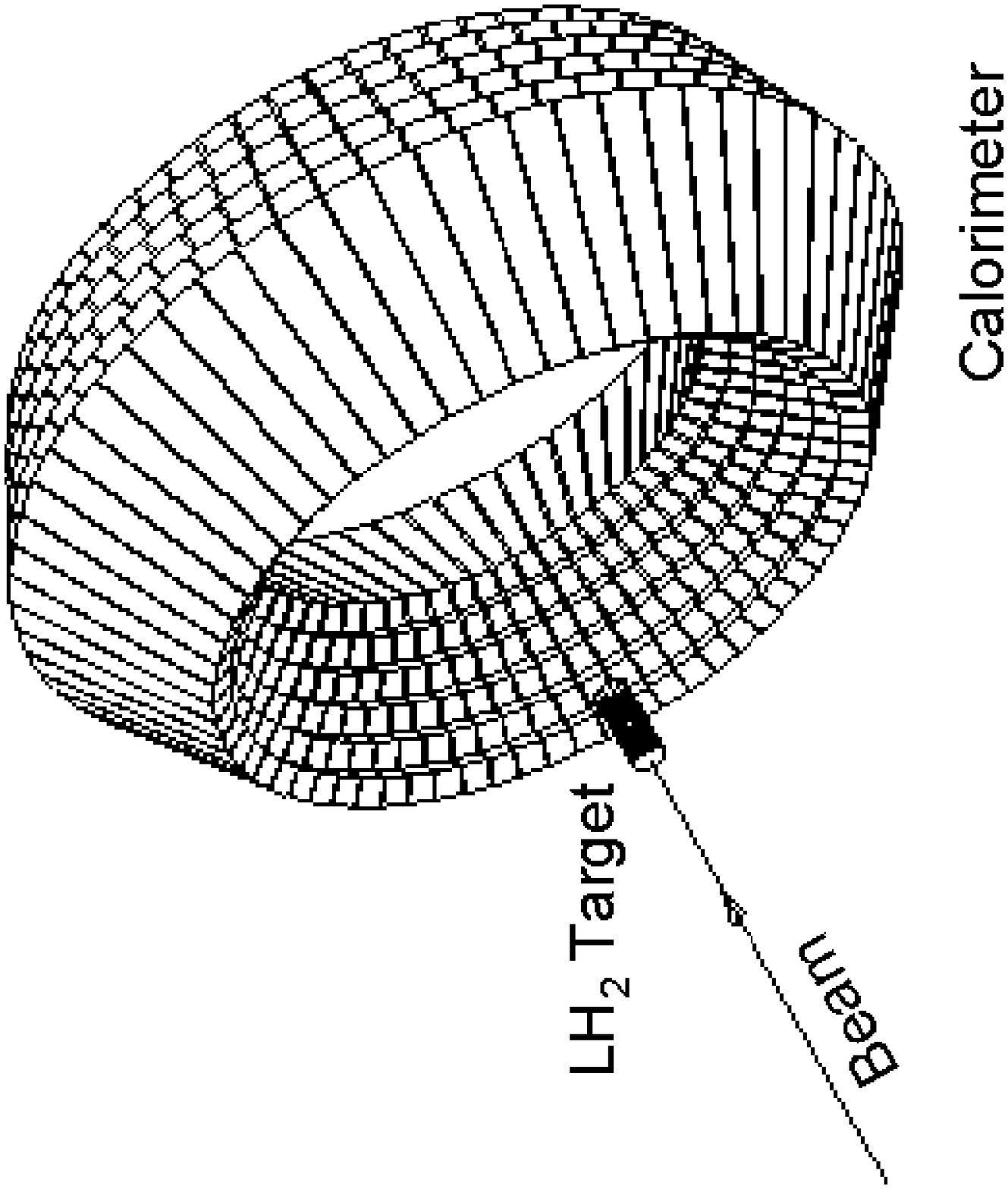}
\vspace*{13pt}
\fcaption{Schematic of the parity-violating electron scattering experimenta PVA4 
being performed at the Mainz Microtron.  The calorimeter consists of an array of 
1022 PbF$_2$ crystals used to count elastically scattered electrons.}
\label{fig:PVA4}
\end{figure}

The PVA4 experiment will initially utilize the 855 MeV beam from the
MAMI accelerator to measure the parity-violating elastic scattering
asymmetry at an angle centered around $35^\circ$ ($Q^2=0.23$ GeV$^2$).
This forward angle asymmetry will yield a measurement of the quantity
$G_E^s + 0.21 G_M^s$; the measurement began in summer
2000.  A 20 $\mu$A beam with 80\% polarization is incident on a
10 cm LH$_2$ target for the experiment.  The detector for the
experiment consists of 1022 PbF$_2$ calorimeter crystals covering
a solid angle of 0.7 sr and arranged in a pointing geometry relative
to the target as shown in Figure~\ref{fig:PVA4}.  The first
measurements will be made with half the detectors arranged in two
diametrically opposed quarters covering half the total azimuthal
angle.  The fast Cerenkov signal from the PbF$_2$ allows separation of
elastic and inelastic electrons in hardware.  Using an analog sum of
signals from a central detector and its eight nearest neighbors an
energy resolution of about 3.5\% has been achieved with an integration
gate of 20 ns.  This allows effective separation of elastic and
inelastic electrons -- the inelastic yield being about x10 larger than
that from elastic scattering.  The same apparatus can be reversed
relative to the beam to provide corresponding asymmetries over a range
of momentum transfers at backward angles (i.e. with a scattering
angle of $145^\circ$).

\begin{figure}[htbp]
\vspace*{13pt}
\vspace*{3.0truein}		
\includegraphics{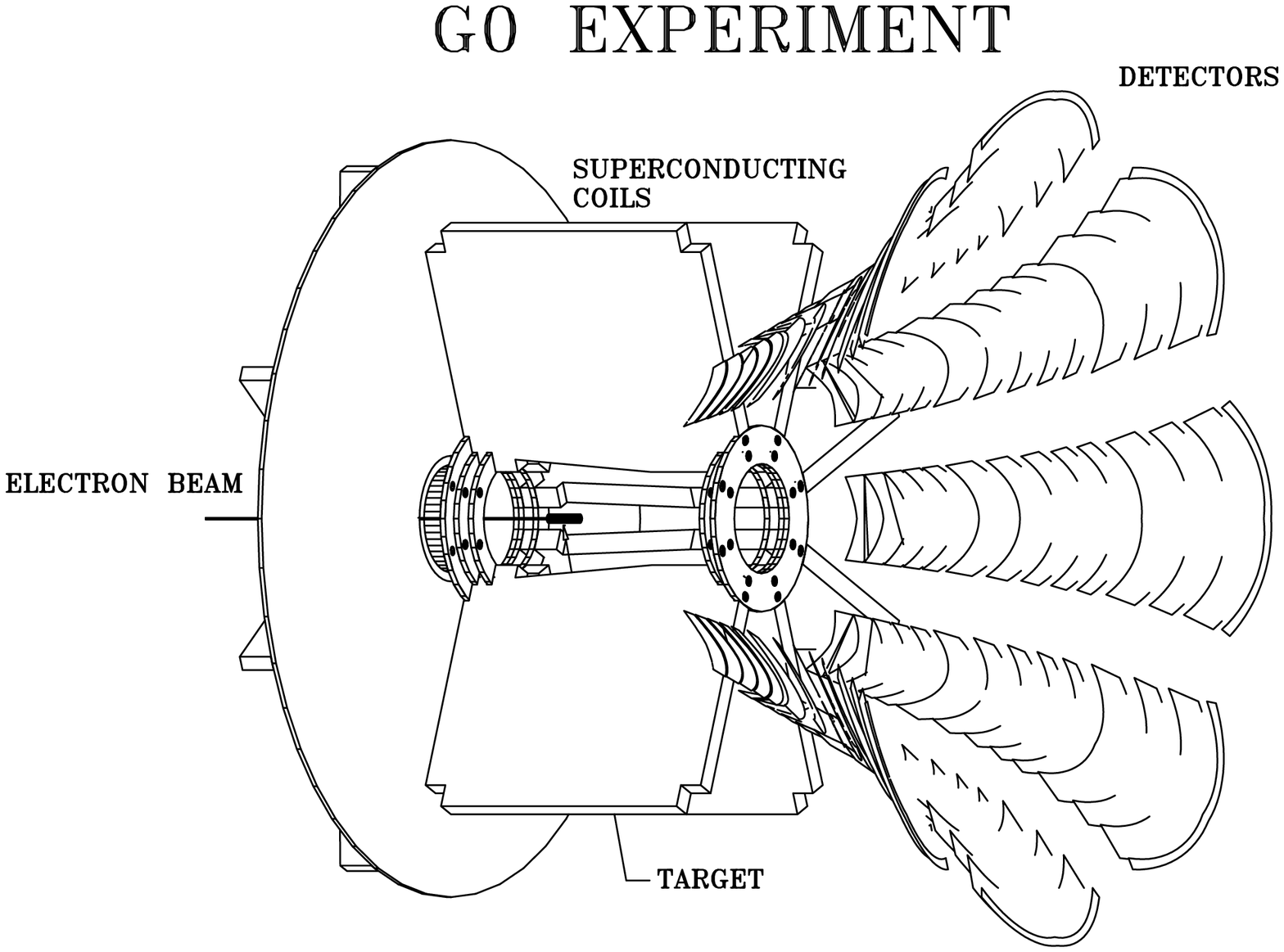}
\vspace*{13pt}
\fcaption{Schematic of the G0 parity-violating electron scattering experiment 
to 
be performed at JLab.  A dedicated superconducting toroidal 
spectrometer will be used to detect recoil protons for forward angle 
measurements and electrons for back angle measurements.}
\label{fig:G0}
\end{figure}

The goal of the G0 experiment is to measure forward proton asymmetries
and backward asymmetries for both the proton and deuteron in order to provide a
complete set of observables from which the charge, magnetic and axial
neutral weak currents of the nucleon can be determined.  It will
utilize a 40 $\mu$A, 70\% polarized beam from the JLab accelerator.
The experimental apparatus consists of a superconducting toroidal
magnet used to focus particles from a 20 cm liquid hydrogen target to
an array of plastic scintillator pairs located outside the magnet cryostat
(see Figure~\ref{fig:G0}).

In the first G0 experiment, forward angle asymmetries will be measured
by detecting the recoil protons from elastic scattering.  With an
acceptance of about 0.9 sr (for scattering angles centered at about
$70^\circ$), the spectrometer will measure asymmetries over the range
$0.12 < Q^2 < 1.0$ GeV$^2$ with a beam energy of 3 GeV.  For this
measurement elastic protons are identified by time-of-flight
(discriminating against inelastic protons and faster $\beta\sim 1$
particles such as $\pi^+$) and their $Q^2$ is determined by where in
the focal surface they are detected.  Custom electronics is being used
for fast accumulation of t.o.f. spectra with resolutions of 0.25 - 1
ns -- the maximum elastic rate in the scintillator pairs is about 1
MHz.  Background yields and asymmetries are thus measured concurrently
and will be used to correct the elastic asymmetries.

Backward angle asymmetries will be measured with the same apparatus,
by reversing it relative to the beam direction.  In this case
elastically scattered electrons will be measured at scattering angles
around $110^\circ$.  A set of smaller scintillators will be installed
near the exit window of the cryostat to discriminate elastic and
inelastic electrons.  In combination with the scintillators in the
focal surface this allows a rough measurement of both electron
momentum and scattering angle -- elastic electrons will appear only in
certain well defined pairs of detectors.  Measurements of
quasi-elastic scattering from deuterium at backward angles will
require improved particle i.d. to separate electrons and $\pi^-$ (essentially
absent in the hydrogen measurements).  In principle, the G0 experiment
will therefore also be able to investigate the effective axial current seen by
the electron over the full range of momentum transfers of the experiment.

\section{Summary}

The quark and gluon sea of the nucleon is, particularly at low
energies, an important, relatively unknown part of its structure.
Measurements of strange quark matrix elements of the nucleon represent
direct windows on at least part of this structure, to the extent that
strange and light quarks in the sea have some features in common.
Parity-violating electron scattering provides a new, relatively clean
determination of the contributions of the three lightest flavors to
nucleon vector currents (the ordinary charge and magnetic form
factors).

The asymmetries measured in these experiments determine the
interference between the weak neutral and electromagnetic nucleon
currents.  The weak neutral current, in turn, is related to the same
matrix elements as the electromagnetic current, but is weighted by the
weak charges given by the Standard Model (assuming point-like, spin
1/2 quarks).  Thus, by measuring the parity-violating scattering and
making one further assumption that the proton and neutron obey charge
symmetry, three observables can be used to determine the contributions
of the three lightest flavors, in particular those of the strange
quarks.

Given the difficulty of including dynamical quarks in lattice QCD
calculations, one must presently rely on models to calculate strange quark
matrix elements.  A number of different, but more or less standard
approaches have been made including those associated with chiral
perturbation theory, with various hadronic bases (including simple
loop diagrams, vector dominance or dispersion relations), with
approximations of QCD (Skyrme and Nambu-Jona-Lasinio) as well as with
somewhat more microscopic constitutent quark models.  There is some
consistency among the models, which generally indicate a negative
contribution to the magnetic moment and a small strangeness charge radius.
However, the agreement of these results with present experiments is
not obvious, and in any case the trend of recent
investigations is to show that any such predictions are typically very
sensitive to some of the assumptions in the calculations.  This likely
reflects the fact that strange quark matrix elements are rather
detailed aspects of a structure whose main features are even difficult to
model at this stage.  The key question is here is whether any present
theoretical scheme can reliably estimate subtle dynamic effects such as the
strange matrix elements of the nucleon.

Strange quark vector currents are now being extracted from
parity-violation measurements.  Even though the asymmetries to be
measured are very small -- on the scale of parts-per-million -- the
remarkable precision of polarized electron beams makes such
experiments more or less routine.  Indeed, in the two completed
experiments, the false asymmetries due to helicity-correlated beam
changes are essentially negligible compared to other systematic
uncertainties.  The results from the SAMPLE and HAPPEX experiments
suggest small contributions of strange quarks to nucleon matrix
elements, perhaps on the scale of 0 - 10\% of the total.  One of the
important lessons from the first experiments is that other nucleon
form factors are important in extracting this information.  The
results are, for example, quite sensitive to the neutron magnetic form 
factor.  In addition, the SAMPLE measurement on deuterium indicates
that the axial current measured in these experiments has significant
differences from that measured in neutrino scattering and that
calculations of these effects (related to the nucleon anapole moment)
may not be reliable.

Both experiment and theory are at an early stage in this formidable
problem.  The next set of experiments is poised to broaden the range
of momentum transfer and to separate the charge and magnetic
components of the currents.  Particularly because we are at present
trying to bootstrap ourselves to a clearer picture of the structure,
the basic observables must be measured.  Such measurements will
provide an important foundation for detailed understanding of
forthcoming lattice calculations and their relation to the familiar
nucleon models.

\begin{center}
{\bf Acknowledgement}
\end{center}

It is a pleasure to acknowledge useful input from M.J. Ramsey-Musolf
and H. Forkel.  This work was supported in part by the National
Science Foundation.

\end{document}